 \documentclass[final,5p,times,twocolumn]{elsarticle}

\usepackage{amsmath,amsfonts,amssymb,mathtools}
\usepackage{bbm}
\usepackage{graphicx}
\usepackage{exscale}
\usepackage{hhline}

\newtheorem{thm}{Theorem}

\newtheorem{rem}[thm]{Remark}
\newtheorem{exmp}{Example}

\newcommand{\E}[1]{\mathop{{\rm \bf E}\left\{#1\right\}}\nolimits}
\DeclareMathAlphabet\mathbfcal{OMS}{cmsy}{b}{n}

\usepackage{clrscode}

\begin{document}

\begin{frontmatter}
\title{Factored-form Kalman-like Implementations under Maximum Correntropy Criterion}

\author[CEMAT]{Maria V.~Kulikova\corref{cor}} \ead{maria.kulikova@ist.utl.pt} \cortext[cor]{Corresponding
author.}

\address[CEMAT]{CEMAT (Center for Computational and Stochastic Mathematics), Instituto Superior T\'{e}cnico, Universidade de Lisboa, \\ Av. Rovisco Pais 1,  1049-001 Lisboa, Portugal}

\begin{abstract}
The maximum correntropy criterion (MCC) methodology is recognized to be a robust filtering strategy with respect to outliers and shown to outperform the classical Kalman filter (KF) for estimation accuracy in the presence of non-Gaussian noise. However, the numerical stability of the newly proposed MCC-KF estimators
in finite precision arithmetic is seldom addressed. In this paper, a family of {\it factored-form} (square-root) algorithms is derived for the MCC-KF and
 its improved variant, respectively. The family traditionally consists of three factored-form implementations:  (i) Cholesky factorization-based algorithms, (ii) modified Cholesky, i.e. UD-based methods, and (iii) the recently established SVD-based filtering. All these strategies are commonly recognized to enhance
the numerical robustness of conventional filtering with respect to roundoff errors and,  hence, they are the preferred implementations when solving applications with high reliability requirements. Previously, only Cholesky-based IMCC-KF algorithms have been designed. This paper enriches a factored-form family by introducing the UD- and SVD-based methods as well. A special attention is paid to {\it array} algorithms that are proved to be the most numerically stable and, additionally, suitable for parallel implementations. The theoretical properties are discussed and numerical comparison is presented for determining the most reliable implementations.
\end{abstract}


\end{frontmatter}

\section{Introduction and problem statement}

Consider a linear discrete-time stochastic system
  \begin{align}
   x_{k} = & F_{k-1} x_{k-1}  + G_{k-1} w_{k-1},  & k & \ge 1 \label{eq:st:1} \\
    y_k  = & H_k x_k + v_k   \label{eq:st:2}
  \end{align}
where $F_k \in \mathbb R^{n\times n}$, $G_k \in \mathbb R^{n\times q}$, $H_k \in \mathbb R^{m\times n}$. The vectors $x_k \in \mathbb R^n$  and $y_k \in \mathbb R^m$ are the unknown dynamic state and the available measurements, respectively. The random variables $\{ x_0, w_k, v_k\}$ have the following properties:
$$
\E{
\begin{bmatrix}
 x_0 \\ w_k\\v_{k}
\end{bmatrix}
\begin{bmatrix}
x_0^T & w_j^T&v_{j}^T & 1
\end{bmatrix}}=
\begin{bmatrix}
\Pi_0 & 0 & 0 & \bar x_0 \\
0 & Q_k\delta_{kj} &  0 & 0  \\
0 & & R_k\delta_{kj}  &  0
\end{bmatrix}
$$
where $Q_k \in \mathbb R^{q\times q}$, $R_k \in \mathbb R^{m\times m}$ and $\delta_{kj}$ denotes the Kronecker delta function.

The goal of any filtering method is to recover the unknown random sequence $\{ x_{k} \}_1^N$ from the observed one $\{ y_k\}_1^N$. The classical Kalman filter (KF) yields the minimum {\it linear} expected mean square error (MSE) estimate $\hat x_{k|k}$  of the state vector $x_k$, given measurements $\mathcal{Y}_1^k = \{y_1, \ldots, y_k \}$, i.e.
\begin{equation}
\mbox{arg}\!\!\min \limits_{\!\!\!\!\!\!\!\!\!\!\hat x_{k|k} \in span({\cal Y}_1^k)} \E{\|x_k - \hat x_{k|k} \|^2}.   \label{eq:LMSE}
\end{equation}

If the examined state-space model is Gaussian, i.e. $w_k$, $v_k$ and the initial state $x_0$  are jointly Gaussian, then estimator~\eqref{eq:LMSE} coincides with the  minimum expected MSE estimator~\cite{2017:Aravkin}.
\begin{equation}
\mbox{arg}\min \limits_{\!\!\!\hat x_{k|k}} \E{ \|x_{k} - \hat x_{k|k} \|^2}.  \label{eq:MSE}
\end{equation}

In other words, although the classical KF is a linear estimator, in Gaussian settings it yields the minimum
expected MSE estimate. In general (non-Gaussian case), the classical KF produces only sub-optimal solution for estimation problem~\eqref{eq:MSE}.
To deal with non-Gaussian uncertainties and outliers/impulsive noise in state-space model~\eqref{eq:st:1}, \eqref{eq:st:2}, various ``distributional robust'' filtering/smoothing methods have been developed in engineering literature. For detecting outliers, the Huber-based and M-estimator-based KF algorithms suggest to construct weight matrices (or scalars) and utilize them for inflating the innovation or measurement noise covariances for reducing the estimation error~\cite{1977:Masreliez,2010:Hajiyev,2013:Chang}. The unknown input filtering (UIF) methodology suggests to model unknown external excitations as unknown inputs and, next, to derive the robust observer~\cite{2012:Charandabi,2014:Charandabi}. Meanwhile, the most recent and comprehensive survey of existed Kalman-like smoothing methods developed for non-Gaussian state-space models can be found in~\cite{2017:Aravkin}. In this paper, an alternative strategy called the maximum correntropy criterion (MCC) filtering is in the focus. It becomes an important topic for analysis in the past few years, both for linear~\cite{2007:Liu,2012:Cinar,2014:Chen,2015:Chen,2016:Izanloo,2017:Chen} and nonlinear systems~\cite{2018:Kulikov:SP,2017:Liu:UKF,2017:Qin}.

The correntropy represents a similarity measure of two random variables. It can be used as an optimization cost in related estimation problem as discussed in~\cite[Chapter~5]{2018:Principe:book}: an estimator of unknown state $X \in {\mathbb R}$ can be defined as a function of observations $Y \in {\mathbb R}^m$, i.e. $\hat X = g(Y)$ where $g$ is solved by maximizing the correntropy between $X$ and $\hat X$ that is~\cite{2012:Chen}
\begin{equation} \label{mcc:kriterion}
g_{MCC} = \mbox{arg}\max \limits_{g \in G} V(X,\hat X) = \mbox{arg}\max \limits_{g \in G} \E{k_{\sigma}\Bigl(X - g(Y)\Bigr)}
\end{equation}
where $G$ stands for the collection of all measurable functions of $Y$, $k_{\sigma}(\cdot)$ is a kernel function and  $\sigma > 0$ is the kernel size (bandwidth). One of the most popular kernel function utilized in practice is the Gaussian kernel given as follows:
\begin{equation}\label{Gauss_kernel}
k_{\sigma}(X - \hat X) = \exp \left\{ -{(X - \hat X)^2}/{(2\sigma^2)}\right\}.
\end{equation}
It is not difficult to see that the MCC cost~\eqref{mcc:kriterion} with Gaussian kernel~\eqref{Gauss_kernel} reaches its maximum if and only if $X = \hat X$.

In~\cite{2012:Cinar,2016:Izanloo}, the MCC estimation problem~\eqref{mcc:kriterion} with kernel~\eqref{Gauss_kernel} is combined with the
 minimum {\it linear} expected MSE estimation problem related to the classical KF in~\eqref{eq:LMSE}. The resulted estimator is called the MCC-KF method. Taking into account that only a finite number of data points $k=1, \ldots N$ is available in practice, the sample mean is utilized in corresponding formulas. The problem of estimating $x_k$ for state-space model~\eqref{eq:st:1}, \eqref{eq:st:2} is equivalent to maximizing $J(k)$ given by
\begin{equation}
\hat x_{k|k}  = \mbox{arg}\max J(k)
\end{equation}
where
\begin{equation}
 J(k)  = k_{\sigma}(\|\hat x_{k|k}-F_{k-1}\hat x_{k-1|k-1}\|_{P_{k|k-1}^{-1}})  + k_{\sigma}(\|y_k-H_k\hat x_{k|k}\|_{R_k^{-1}})
\end{equation}
with the Gaussian kernel functions defined as follows~\cite{2016:Izanloo}:
\begin{align*}
k_{\sigma}(\|\hat x_{k|k}\!-F_{k-1}\hat x_{k-1|k-1}\|_{P_{k|k-1}^{-1}}\!) &\! = \exp \Bigl\{ -\frac{\|\hat x_{k|k}-F_{k-1}\hat x_{k-1|k-1}\|_{P_{k|k-1}^{-1}}^2}{2\sigma^2}\Bigr\}, \\
k_{\sigma}(\|y_k-H_k\hat x_{k|k}\|_{R_k^{-1}}) &\! = \exp \Bigl\{ -\frac{\|y_k-H_k\hat x_{k|k}\|_{R_k^{-1}}^2}{2\sigma^2}\Bigr\}.
\end{align*}

The optimization condition $\partial J(k)/ \partial  \hat x_{k|k} = 0$ yields the nonlinear equation that should be solved with respect to $\hat x_{k|k}$
\begin{align}
\hat x_{k|k} & =  F_{k-1} \hat x_{k-1|k-1} \nonumber \\
& + \frac{k_{\sigma}\left(\|y_k-H_k \hat x_{k|k}\|_{R_k^{-1}}\right) }{k_{\sigma}\left(\|\hat x_{k|k} -F_{k-1}\hat x_{k-1|k-1}\|_{P_{k|k-1}^{-1}}\right)}H_k^T\!(y_k - H_k \hat x_{k|k}). \label{eq:solv:1}
\end{align}

In~\cite{2012:Cinar,2016:Izanloo}, a fixed point rule (with one iterate) is utilized for solving~\eqref{eq:solv:1} where the initial approximation $\hat x_{k|k}^{(0)}$ is set to $\hat x_{k|k-1}$, i.e. $\hat x_{k|k}^{(0)} = \hat x_{k|k-1}$ is substituted at the right-hand side of equation~\eqref{eq:solv:1}. Thus, we get
\begin{equation}
\hat x_{k|k}  =  F_{k-1}\hat x_{k-1|k-1}  + \lambda_k H_k^T(y_k - H_k \hat x_{k|k-1}) \label{eq:result}
\end{equation}
where $\lambda_k$ stands for
\begin{equation}
\lambda_{k}  = \frac{k_{\sigma}(\|y_k-H_k \hat x_{k|k-1}\|_{R^{-1}_k})}{k_{\sigma}(\|\hat x_{k|k-1}-F_{k-1} \hat x_{k-1|k-1}\|_{P_{k|k-1}^{-1}})}.  \label{eq:lambda}
\end{equation}

Finally, the recursion for the state estimate in~\eqref{eq:result} is utilized with the KF-like estimation and the related error covariance propagation~\cite{2016:Izanloo}. The resulted estimator is called the maximum correntropy criterion Kalman filter (MCC-KF) and summarized as follows~\cite[p.~503]{2016:Izanloo}. For readers' convenience, it is presented here in the form of Algorithm~1.
\begin{codebox}
\Procname{{\bf Algorithm 1}. $\proc{MCC-KF}$ ({\it original MCC-KF})}
\zi \textsc{Initialization:}($k=0$) $\hat x_{0|0} = \bar x_0$ and $P_{0|0} = \Pi_0$.
\zi \textsc{Time Update}: ($k=\overline{1,N}$)
\li \>$\hat x_{k|k-1}  = F_{k-1} \hat x_{k-1|k-1}$; \label{mcc:p:X}
\li \>$P_{k|k-1}  = F_{k-1} P_{k-1|k-1}F_{k-1}^T+G_{k-1}Q_{k-1}G_{k-1}^T$; \label{mcc:p:P}
\zi \textsc{Measurement Update}: ($k=\overline{1,N}$)
\li \>Compute $\lambda_k$ by formula~\eqref{eq:lambda};  \label{mcc:f:L}
\li \>$K_{k}  = \lambda_k \left( P_{k|k-1}^{-1}+\lambda_k H_k^T R_k^{-1} H_k \right)^{-1}H_k^T R_k^{-1}$; \label{mcc:f:K}
\li \>$P_{k|k}  = (I - K_{k}H_k)P_{k|k-1}(I - K_{k}H_k)^T+K_k R_k K_k^T$;  \label{mcc:f:P}
\li \>$\hat x_{k|k}  =    \hat x_{k|k-1}+K_{k}(y_k-H_k \hat x_{k|k-1})$.   \label{mcc:f:X}
\end{codebox}

It is worth noting here that the MCC-KF coincides with the classical KF when $\lambda_k=1$. Thus, similar to the classical KF equations presented in~\cite[p.~128-129]{simon2006optimal}, the following formulas have been obtained for the MCC-KF method in~\cite[Lemma~1]{2017:CSL:Kulikova}:
\begin{align}
K_{k} & = \lambda_k P_{k|k-1}H_k^T\left(\lambda_kH_kP_{k|k-1}H_k^T+R_k\right)^{-1} \label{mcc:K:eq1}  \\
& = \lambda_k P_{k|k}H_k^TR_k^{-1}  \label{mcc:K:eq2}
\end{align}
where $R_{e,k} := \lambda_kH_kP_{k|k-1}H_k^T+R_k$, and $P_{k|k}$ satisfies
\begin{align}
P_{k|k}  & = \left( P_{k|k-1}^{-1}+\lambda_kH_k^TR_k^{-1}H_k\right)^{-1} \label{mcc:P:eq1} \\
 &  = (I - K_{k}H_k)P_{k|k-1} \label{mcc:P:eq2} \\
 &  = (I - K_{k}H_k)P_{k|k-1}(I  -  \lambda_k K_{k}H_k)^T+K_kR_kK_k^T. \label{mcc:P:eq3}
\end{align}

It is not difficult to see that the gain matrix $K_k$ in the MCC-KF implementation (see line~\ref{mcc:f:K} in Algorithm~1) is computed by formula~\eqref{mcc:K:eq2} where the error covariance matrix $P_{k|k}$ obeys equation~\eqref{mcc:P:eq1}, i.e. the following formula  holds: $P_{k|k} = \lambda_k \left( P_{k|k-1}^{-1}+\lambda_kH_k^TR_k^{-1}H_k\right)^{-1}H_k^TR_k^{-1}$. Next, having compared equation~\eqref{mcc:P:eq3} for computing $P_{k|k}$ with the  equation in line~\ref{mcc:f:P} of Algorithm~1, we conclude that the suggested MCC-KF implementation (Algorithm~1) neglects the scalar parameter $\lambda_k$ in~\eqref{mcc:P:eq3} in order to keep the symmetric form. In the KF community, the symmetric equation (see line~\ref{mcc:f:P} of Algorithm~1) is called the Joseph stabilized form. It is recognized to be the preferable implementation strategy for the classical KF, because it ensures the symmetric form of error covariance matrix $P_{k|k}$ in the presence of roundoff errors and, hence, improves the numerical robustness~\cite{simon2006optimal,GrewalAndrews2015}. In summary, the MCC-KF implies the simplified error covariance computation strategy ($\lambda_k$ is omitted) in order to keep the reliable Joseph stabilized form. However, a loss in estimation quality is the price to be paid; see the numerical results and the improved MCC-KF (IMCC-KF) variant developed in~\cite{2017:CSL:Kulikova}.  For readers' convenience, the IMCC-KF is summarized in Algorithm~2.

\begin{codebox}
\Procname{{\bf Algorithm 2}. $\proc{IMCC-KF}$ ({\it improved MCC-KF})}
\zi \textsc{Initialization:}($k=0$) $\hat x_{0|0} = \bar x_0$ and $P_{0|0} = \Pi_0$.
\zi \textsc{Time Update}: ($k=\overline{1,N}$)
\li \>$\hat x_{k|k-1}  = F_{k-1} \hat x_{k-1|k-1}$; \label{imcc:p:X}
\li \>$P_{k|k-1}  = F_{k-1} P_{k-1|k-1}F_{k-1}^T+G_{k-1}Q_{k-1}G_{k-1}^T$; \label{imcc:p:P}
\zi \textsc{Measurement Update}: ($k=\overline{1,N}$)
\li \>Compute $\lambda_k$ by formula~\eqref{eq:lambda};  \label{imcc:f:L}
\li \>$K_{k}  = \lambda_k P_{k|k-1}H_k^T\left(\lambda_kH_kP_{k|k-1}H_k^T+R_k\right)^{-1}$; \label{imcc:f:K}
\li \>$P_{k|k}  = (I - K_{k}H_k)P_{k|k-1}$;  \label{imcc:f:P}
\li \>$\hat x_{k|k}  =    \hat x_{k|k-1}+K_{k}(y_k-H_k \hat x_{k|k-1})$.   \label{imcc:f:X}
\end{codebox}

As can be seen, the IMCC-KF (Algorithm~2) implies equation~\eqref{mcc:K:eq1} for the gain matrix $K_k$ computation (line~\ref{imcc:f:K} in Algorithm~2) and formula~\eqref{mcc:P:eq2} for the error covariance matrix $P_{k|k}$ calculation (line~\ref{imcc:f:P} in Algorithm~2).
 Both the MCC-KF and IMCC-KF are shown to outperform the classical KF for estimation accuracy in case of non-Gaussian uncertainties and outliers~\cite{2012:Cinar,2016:Izanloo,2017:CSL:Kulikova}.  Meanwhile, to deal with the numerical instability problem and robustness with respect to roundoff errors, we derive the so-called {\it factored-form} implementations~\cite{2010:Grewal:IEEE}: ``It was recognized in the KF community that the factored-form (square-root) algorithms are the preferred implementations when a high operational reliability is required''.

The key idea of the factored-form methodology for implementing linear and nonlinear KF-like estimators is to factorize the error covariance matrix $P$ in the
form of $P = SS^T$ and, then, re-formulate the underlying filtering equations in terms of these factors, only. Thus, the square-root algorithms recursively update the factors $S$ instead of entire matrix $P$ at each filtering step. It ensures the theoretical properties of any covariance matrix, i.e. its symmetric form and positive semi-definiteness, in the presence of roundoff errors. Indeed, when all measurements are processed, the full matrix $P$ is re-constructed from the updated factors $S$ by backward multiplication $SS^T=P$. Although, the roundoff errors influence the factors $S$, the product $SS^T=P$ is a  symmetric and positive semi-definite matrix.

It is worth noting here that the factorization $P = SS^T$ can be implemented in various ways. This yields a variety of the factored-form (square-root) strategies: (i) Cholesky factorization-based algorithms, e.g. in~\cite{Morf1975,Sayed1994,1995:ParkKailath}; (ii) UD-based implementations in~\cite{Bierman1977}, and (iii) SVD-based methods published recently in~\cite{2017:IET:KulikovaTsyganova}. We stress that all cited square-root methods have been derived for the classical KF. Meanwhile for the MCC methodology the Cholesky-based IMCC-KF implementation has been proposed, only~\cite{2017:CSL:Kulikova}. The goal of this paper is to proceed our recent research and suggest a complete factored-form family for both the MCC-KF (Algorithm~1) and IMCC-KF (Algorithm~2) estimators. Finally, it is important for further derivation that the scalar $\lambda_k$ in the MCC-KF and IMCC-KF methods is a nonnegative value and, hence, a square root exists (for real nonnegative numbers).

\section{The Cholesky factored-form implementations}  \label{SR:filters}

The Cholesky factorization-based approach is the most popular strategy for designing factored-form implementations and, hence, traditionally used in engineering literature. It implies factorization of a symmetric positive definite matrix $A$ in the form $A=(A^{1/2})(A^{1/2})^T$ where the factor $A^{1/2}$ is an upper or lower triangular matrix with positive diagonal elements. The resulted filtering methods belong to factored-form (square-root) family because the matrix square root $S$ (i.e. $P=SS^T$) can be defined as $S:=P^{1/2}$. All algorithms derived in this paper utilize the Cholesky decomposition in the form ${A=A^{T/2} A^{1/2}}$ where $A^{1/2}$ is an upper triangular matrix with positive diagonal elements\footnote{Notation to be used: $A^{T/2} \equiv  (A^{1/2})^T$, ${A^{-1/2} \equiv (A^{1/2})^{-1}}$, ${A^{-T/2} \equiv (A^{-1/2})^T}$.}.

Previously, the {\it array} Cholesky-based implementations have been developed for the IMCC-KF (Algorithm~2), only~\cite{2017:CSL:Kulikova}. Thus, they are not  presented here in details, but their key properties are discussed. The so-called {\it array} form is the preferable filtering implementation because of the following reasons~\cite{1995:ParkKailath}: (i) it makes algorithms convenient for practical use and suitable for parallel implementation; (ii) utilization of stable orthogonal transformation at each iteration step improves numerical stability. In summary, the array Cholesky-based filters utilize $QR$ factorization for updating the corresponding Cholesky factors as follows: the filter quantities are compiled into the pre-array ${\mathbb A}$ and, next, an orthogonal operator $\mathfrak{V}$ is applied $\mathfrak{V}{\mathbb A}={\mathbb R}$ in order to obtain the required triangular form of the post-array ${\mathbb R}$. The updated filter quantities are simply read-off from the post-array ${\mathbb R}$. To summarize, the Cholesky-based IMCC-KF algorithm implies the following factorizations for updating $P_{k|k-1}^{1/2}$ and $P_{k|k}^{1/2}$ at the time and measurement steps~\cite[Algorithm~2]{2017:CSL:Kulikova}:
\begin{align}
\mathfrak{V}
\underbrace{
\begin{bmatrix}
P_{k-1|k-1}^{1/2}F_{k-1}^T\\
Q_{k-1}^{1/2}G^T_{k-1}
\end{bmatrix}
}_{\mbox{\scriptsize Pre-array } {\mathbb A}}
  & =
\underbrace{
\begin{bmatrix}
P_{k|k-1}^{1/2} \\
0
\end{bmatrix}
}_{\mbox{\scriptsize Post-array } {\mathbb R}}  \label{array:1} \\
\mathfrak{W}
\underbrace{
\begin{bmatrix}
R_k^{1/2} & 0  \\
\lambda_k^{1/2}P_{k|k-1}^{1/2}H_k^T & P_{k|k-1}^{1/2}
\end{bmatrix}
}_{\mbox{\scriptsize Pre-array } {\mathbb A}}
  & =
\underbrace{
\begin{bmatrix}
 R_{e,k}^{1/2} & \bar K_{k}^T \\
 0 & P_{k|k}^{1/2}
\end{bmatrix}
}_{\mbox{\scriptsize Post-array } {\mathbb R}} \label{array:2}
\end{align}
where $\bar K_k = \lambda_k^{-1/2} K_k R_{e,k}^{T/2} = \lambda_k^{1/2}P_{k|k-1}H_k^TR_{e,k}^{-1/2}$ is the ``normalized'' feedback gain and $R_{e,k} = \lambda_kH_kP_{k|k-1}H_k^T+R_k$.

Formulas~\eqref{array:1}, \eqref{array:2} are algebraic equivalent to corresponding equations in the conventional IMCC-KF implementation (Algorithm~2). It is not difficult to prove, if we note that the orthogonal transformations set up a conformal (i.e. a norm- and angle-preserving) mapping between the (block) columns of the pre-array ${\mathbb A}$ and the columns of the post-array ${\mathbb R}$. More precisely, let us consider equation~\eqref{array:2} in detail. Because of a norm-preserving mapping, the first inner product is
\[
<\![R_k^{1/2} \quad \!\! \lambda_k^{1/2}P_{k|k-1}^{1/2}H_k^T],[R_k^{1/2} \quad \!\! \lambda_k^{1/2}P_{k|k-1}^{1/2}H_k^T]\!>
= <\![X \quad \!\!0],[X \quad \!\! 0]\!>\!.
\]
Hence, we get $X^TX = \lambda_kH_k\underbrace{P_{k|k-1}^{T/2}P_{k|k-1}^{1/2}}_{P_{k|k-1}}H_k^T+\underbrace{R_k^{T/2}R_k^{1/2}}_{R_k} = R_{e,k}$, i.e. $X:= R_{e,k}^{1/2}$. At the same way, we define
\begin{align*}
<\![R_k^{1/2} \quad \!\! \lambda_k^{1/2}P_{k|k-1}^{1/2}H_k^T],[0 \quad \!\! P_{k|k-1}^{1/2}]\!>
& = <\![R_{e,k}^{1/2} \quad \!\!0],[Y \quad \!\! Z]\!>\!, \\
<\![0 \quad \!\! P_{k|k-1}^{1/2}],[0 \quad \!\! P_{k|k-1}^{1/2}]\!>
& = <\![Y \quad \!\! Z],[Y \quad \!\! Z]\!>\!
\end{align*}
and obtain the set of equations
\begin{align*}
\lambda_k H_k P_{k|k-1}& = R_{e,k}^{T/2}Y, &  P_{k|k-1} & = Z^TZ + Y^TY.
\end{align*}
Thus, we get $Y := \lambda_k^{1/2} R_{e,k}^{-T/2} H_k P_{k|k-1} = \bar K_k^T$ and, hence,
\begin{align*}
Z^TZ & = P_{k|k-1}- Y^TY = P_{k|k-1}- \bar K_k \bar K_k^T \\
& = P_{k|k-1} - \lambda_kP_{k|k-1}H_k^TR_{e,k}^{-1}H_kP_{k|k-1} \\
& = P_{k|k-1} - K_kH_kP_{k|k-1} = (I - K_{k}H_k)P_{k|k-1}=P_{k|k},
\end{align*}
i.e. from~\eqref{mcc:P:eq2} we conclude $P_{k|k} = Z^TZ$ and, hence, $Z:= P_{k|k}^{1/2}$.

Thus, factorization~\eqref{array:2} implies formulas in lines~\ref{imcc:f:K} and~\ref{imcc:f:P} of Algorithm~2. Meanwhile equation~\eqref{array:1} yields formula in line~\ref{imcc:p:P} of Algorithm~2. Indeed,
\begin{align*}
& <\![P_{k-1|k-1}^{1/2}F_{k-1}^T \quad Q_{k-1}^{1/2}G^T_{k-1}],[P_{k-1|k-1}^{1/2}F_{k-1}^T \quad Q_{k-1}^{1/2}G^T_{k-1}]\!> \\
& \qquad = <\![X \quad \!\!0],[X \quad \!\! 0]\!>,
\end{align*}
i.e. we get $X^TX = F_{k-1}P_{k-1|k-1}F_{k-1}^T + G_{k-1}Q_{k-1}G_{k-1}^T$ and, hence, we conclude $X:= P_{k|k-1}^{1/2}$.

The array Cholesky-based MCC-KF implementation (Algorithm~1) can be derived at the same way. Indeed, the time update stage in Algorithms~1 and~2 is the same and, hence, these parts coincide in the Cholesky-based counterparts as well. We conclude that formula~\eqref{array:1} holds for the MCC-KF. The difference is in the measurement update stage. Having analyzed the equation for gain $K_k$ computation in line~\ref{mcc:f:K} of the MCC-KF (Algorithm~1), we conclude that the error covariance matrix is also computed at the same line. Indeed,
$K_{k}  = \lambda_k \left( P_{k|k-1}^{-1}+\lambda_k H_k^T R_k^{-1} H_k \right)^{-1}H_k^T R_k^{-1}$ in the MCC-KF (Algorithm~1) where $(P_{k|k-1}^{-1}+\lambda_k H_k^T R_k^{-1} H_k)^{-1} = P_{k|k}$ according to equation~\eqref{mcc:P:eq1}. However, at the last line of Algorithm~1, matrix $P_{k|k}$ is re-computed by the Joseph stabilized form derived for the classical KF, i.e. with the skipped $\lambda_k$ value; see formula~\eqref{mcc:P:eq3} and the discussion in Section~1. This means that $P_{k|k}$ in the MCC-KF implementations should be always re-computed at the end. Besides, the equations for computing $K_k$ and $P_{k|k}$ in Algorithm~1 should be processed separately. These formulas are incompatible for combining them into unique array (because of the skipped $\lambda_k$), in contrast to the IMCC-KF (Algorithm~2) and equation~\eqref{array:2}. This results into the following implementation.

\begin{codebox}
\Procname{{\bf Algorithm 1a}. $\proc{SR MCC-KF}$ ({\it Cholesky-based MCC-KF})}
\zi \textsc{Initialization:}($k=0$) $\hat x_{0|0} = \bar x_0$ and $P_{0|0}^{1/2} = \Pi_0^{1/2}$ where
\zi \>Cholesky decomposition is applied: $\Pi_0 = \Pi_0^{T/2}\Pi_0^{1/2}$.
\zi \textsc{Time Update}: ($k=\overline{1,N}$)
\li \>$\hat x_{k|k-1}  = F_{k-1} \hat x_{k-1|k-1}$; \label{mcc:sr:p:X}
\li \>Build the pre-array and apply QR factorization:
\zi \>$\mathfrak{V}
\underbrace{
\begin{bmatrix}
P_{k-1|k-1}^{1/2}F_{k-1}^T\\
Q_{k-1}^{1/2}G^T_{k-1}
\end{bmatrix}
}_{\mbox{\scriptsize Pre-array $\mathbb A$}}
   =
\underbrace{
\begin{bmatrix}
P_{k|k-1}^{1/2} \\
0
\end{bmatrix}
}_{\mbox{\scriptsize Post-array $\mathbb R$}} \stackrel{\mbox{\scriptsize read-off}}{\Longrightarrow} \left[P_{k|k-1}^{1/2}\right];$
\zi \textsc{Measurement Update}: ($k=\overline{1,N}$)
\li \>Compute $\lambda_k$ by formula~\eqref{eq:lambda};
\li \>Build the pre-array and apply QR factorization: \label{mcc:sr:f:K}
\zi \>$\mathfrak{W}
\underbrace{
\begin{bmatrix}
P_{k|k-1}^{-T/2}\\
\lambda_k^{1/2}R_{k}^{-T/2}H_{k}
\end{bmatrix}
}_{\mbox{\scriptsize Pre-array $\mathbb A$}}
   =
\underbrace{
\begin{bmatrix}
P_{k|k}^{-T/2} \\
0
\end{bmatrix}
}_{\mbox{\scriptsize Post-array $\mathbb R$}} \stackrel{\mbox{\scriptsize read-off}}{\Longrightarrow} \left[P_{k|k}^{-T/2}\right];$
\li \>Compute $K_{k}  = \lambda_k \left([P_{k|k}^{-T/2}]^T[P_{k|k}^{-T/2}]\right)^{-1}H_k^T R_k^{-1}$;
\li \>$\hat x_{k|k}  =   \hat x_{k|k-1}+K_{k}(y_k-H_k \hat x_{k|k-1})$.
\li \>Build the pre-array and apply QR factorization: \label{mcc:sr:f:P}
\zi \>$\mathfrak{Q}
\underbrace{
\begin{bmatrix}
P_{k|k-1}^{1/2}(I - K_{k}H_k)^T\\
R_{k}^{1/2}K^T_{k}
\end{bmatrix}
}_{\mbox{\scriptsize Pre-array $\mathbb A$}}
   =
\underbrace{
\begin{bmatrix}
P_{k|k}^{1/2} \\
0
\end{bmatrix}
}_{\mbox{\scriptsize Post-array $\mathbb R$}}\stackrel{\mbox{\scriptsize read-off}}{\Longrightarrow} \left[P_{k|k}^{1/2}\right].$
\end{codebox}
\begin{rem}
The values appeared in square brackets in all algorithms in this paper denote the blocks that are directly read-off from the corresponding post-arrays.
\end{rem}

Following our previous analysis, it is not difficult to validate formulas in lines~\ref{mcc:sr:f:K} and~\ref{mcc:sr:f:P} of Algorithm~1a. Indeed, the following set of equations holds:
\begin{align*}
&<[P_{k|k-1}^{-T/2} \quad \!\! \lambda_k^{1/2}R_{k}^{-T/2}H_{k}]><[P_{k|k-1}^{-T/2} \quad \!\! \lambda_k^{1/2}R_{k}^{-T/2}H_{k}]> \\
& \quad = <\![X \quad \!\!0],[X \quad \!\! 0]\!>, \\
&<\![P_{k|k-1}^{1/2}(I - K_{k}H_k)^T \quad \!\! R_{k}^{1/2}K^T_{k}],[P_{k|k-1}^{1/2}(I - K_{k}H_k)^T \quad \!\! R_{k}^{1/2}K^T_{k}]\!> \\
&\quad = <\![Y \quad \!\!0],[Y \quad \!\! 0]\!>\!,
\end{align*}
i.e. $X^TX = P_{k|k-1}^{-1}+\lambda_kH^T_{k}R_{k}^{-1}H_{k}$ and $Y^TY = (I - K_{k}H_k)P_{k|k-1}(I - K_{k}H_k)^T + K_{k}R_{k}K_{k}^T$. Taking into account formula~\eqref{mcc:P:eq1}, we conclude $X:=P_{k|k}^{-T/2}$ and, next, $Y:=P_{k|k}^{1/2}$.

As discussed above, although the inverse $P_{k|k}^{-1/2}$ is already available from line~\ref{mcc:sr:f:K} of Algorithm~1a, we cannot avoid line~\ref{mcc:sr:f:P} for computing $P_{k|k}^{1/2}$ according to the Joseph stabilized equation. In fact, if the second orthogonal transformation at the measurement update step in Algorithm~1a is skipped (i.e. we simply inverse the already computed $P_{k|k}^{-1/2}$ to obtain $P_{k|k}^{1/2}$), then the resulted algorithm is algebraic equivalent to the IMCC-KF (Algorithm~2), but not to the MCC-KF (Algorithm~1). Indeed, the matrix $P_{k|k}^{-1/2}$ calculation in line~\ref{mcc:sr:f:K} is, in fact, formula~\eqref{mcc:P:eq1} that is equivalent to equation~\eqref{mcc:P:eq3}, but not to the symmetric Joseph stabilized form of the classical KF recursion utilized in line~\ref{mcc:sr:f:P} of the MCC-KF (Algorithm~1). Thus, to ensure algebraic equivalence between all MCC-KF implementations, the extra computations related to the symmetric classical KF formula for $P_{k|k}$ update are not avoidable and should be performed in any case.

\section{The UD-based factored-form implementations}  \label{UD:filters}

The first UD-based KF algorithms have been developed by Thornton and Bierman~\cite{1976:Thornton,bierman1977numerical}. The key idea of this strategy is to avoid square-rooting. Due to the computational complexity reasons, the square-root-free methods were preferable for practical implementations in 1970s; see also filtering methods in~\cite{carlson1973fast}. For modern computational devices the square root operation is not a problem because it can be implemented efficiently. However, the UD-based filters still deserve some merit. One of possible reasons is utilization of square-root-free orthogonal rotations that might be more numerically stable than usual QR decomposition; see the discussion in~\cite{Bjorck1967,1991:Gotze,1993:Hsieh,1994:Bjorck}. Thereby, the UD-based estimators' quality and robustness are enhanced in ill-conditioned situations~\cite{2017:IET:KulikovaTsyganova,bierman1977numerical}. It is also worth noting here that the first UD-based KF implementations were derived in sequential form, i.e. when the available measurement vector $y_k$ is processed in a component-wise manner. Nowadays, the advantageous {\it array} form is preferable for practical implementation. For the classical KF, such array implementations have been suggested in~\cite{GrewalAndrews2015,JoverKailathSayed1986} as well as the {\it extended array} algorithms have been derived in~\cite{1991:Chun,2017:Semushin}.  In this section, the array UD-based methods are derived for both the MCC-KF (Algorithm~1) and IMCC-KF (Algorithm~2) estimators.

Consider the modified Cholesky decomposition $P=\bar U_{P}D_{P}\bar U^T_{P}$ where $D_{P}$ denotes a diagonal matrix and $\bar U_{P}$ is an upper triangular matrix with $1$'s on the main diagonal~\cite{Bierman1977}. The UD-based implementations belong to factored-form (square-root) family because
the matrix square root $S$ (i.e. $P=SS^T$) can be defined as $S:=\bar U_{P}D_{P}^{1/2}$. As usual, the underlying filter recursion should be re-formulated for updating the resulted $\bar U_{P}$ and $D_{P}$ factors instead of full matrix $P$. In this paper, the modified weighted Gram-Schmidt (MWGS) orthogonalization is utilized for updating the resulted $UD$ factors as follows~\cite[Lemma~VI.4.1]{Bierman1977}: given ${\mathbb A} \in {\mathbb R}^{r\times s}$, $r \ge s$ and diagonal ${\mathbb D}_{A} \in {\mathbb R}^{r\times r}$ (${\mathbb D}_{A}>0$), compute an unite upper triangular matrix ${\mathbb B} \in {\mathbb R}^{s\times s}$ and diagonal matrix ${\mathbb D}_{B} \in {\mathbb R}^{s\times s}$, i.e.
\begin{equation}
\label{assume:1}
{\mathbb A}= \mathfrak{W}{\mathbb B}^T \quad \mbox{ with } \quad \mathfrak{W}^T{\mathbb D}_{A}\mathfrak{W}={\mathbb D}_{B}
\end{equation}
where $\mathfrak{W}  \in {\mathbb R}^{r\times s}$ is the MWGS orthogonalization.

Taking into account properties of orthogonal matrices, from equation~\eqref{assume:1} we obtain
\begin{equation} {\mathbb A}^T {\mathbb D}_{A} {\mathbb A}  = {\mathbb B}\mathfrak{W}^T {\mathbb D}_{A}\mathfrak{W} {\mathbb B}^T = {\mathbb B} {\mathbb D}_{B} {\mathbb B}^T. \label{general:ud}
\end{equation}

The first equation in~\eqref{assume:1} can be re-written as follows: ${\mathbb A}^T = {\mathbb B}\mathfrak{W}^T$, i.e. the orthogonal transformation sets up a conformal mapping between the (block) rows of the pre-array ${\mathbb A}^T$ and the rows of the post-array ${\mathbb B}$ where a diagonally weighted norms are utilized and preserved. Thus, we derive the following UD-based implementations.

\begin{codebox}
\Procname{{\bf Algorithm 1b}. $\proc{UD MCC-KF}$ ({\it UD-based MCC-KF})}
\zi \textsc{Initialization:}($k=0$) $\hat x_{0|0} = \bar x_0$ and $\bar U_{P_{0|0}} = \bar U_{\Pi_0}$, $D_{P_{0|0}} = D_{\Pi_0}$
\zi \>where UD-decomposition is applied: $\Pi_0 = \bar U_{\Pi_0}D_{\Pi_0}\bar U_{\Pi_0}^T$.
\zi \textsc{Time Update}: ($k=\overline{1,N}$)
\li \>$\hat x_{k|k-1}  = F_{k-1} \hat x_{k-1|k-1}$;
\li \>Build pre-arrays ${\mathbb A}$, ${\mathbb D}_A$ and apply MWGS algorithm: \label{mcc:ud:p:P}
\zi \>$\underbrace{
\begin{bmatrix}
F_{k-1} \bar U_{P_{k-1|k-1}} & \; G_{k-1}\bar U_{Q_{k-1}}
\end{bmatrix}
}_{\mbox{\scriptsize Pre-array ${\mathbb A}^T$}}\!
\! = \!\!
  \underbrace{
\begin{bmatrix}
\bar U_{P_{k|k-1}}
\end{bmatrix}
}_{\mbox{\scriptsize Post-array ${\mathbb B}$}}\!\!{\mathfrak V}^T\stackrel{\mbox{\scriptsize read-off}}{\Longrightarrow} \left[\bar U_{P_{k|k-1}}\right];$
\zi \>$\mathfrak{V}^T\!\!
\underbrace{
\left[{\rm diag}\left\{
D_{P_{k-1|k-1}}, \: D_{Q_{k-1}}
\right\}\right]
}_{\mbox{\scriptsize Pre-array ${\mathbb D}_{A}$}}\!\mathfrak{V}
= \!\! \underbrace{
\begin{bmatrix}
D_{P_{k|k-1}}
\end{bmatrix}
}_{\mbox{\scriptsize Post-array ${\mathbb D}_{B}$}} \stackrel{\mbox{\scriptsize read-off}}{\Longrightarrow} \left[D_{P_{k|k-1}}\right];$
\zi \textsc{Measurement Update}: ($k=\overline{1,N}$)
\li \>Compute $\lambda_k$ by formula~\eqref{eq:lambda};
\li \>Build pre-arrays ${\mathbb A}$, ${\mathbb D}_A$ and apply MWGS algorithm: \label{mcc:ud:f:K}
\zi \>$\underbrace{
\begin{bmatrix}
\bar U_{P_{k|k-1}}^{-T} & \; \lambda_k^{1/2}H_{k}^T\bar U_{R_{k}}^{-T}
\end{bmatrix}
}_{\mbox{\scriptsize Pre-array ${\mathbb A}^T$}}\!
 =\!
  \underbrace{
\begin{bmatrix}
\bar U_{P_{k|k}}^{-T}
\end{bmatrix}
}_{\mbox{\scriptsize Post-array ${\mathbb B}$}}\!\!{\mathfrak W}^T \stackrel{\mbox{\scriptsize read-off}}{\Longrightarrow} \left[\bar U_{P_{k|k}}^{-T}\right];$
\zi \>$\mathfrak{W}^T\!
\underbrace{
\left[{\rm diag}\left\{
D_{P_{k|k-1}}^{-1}, \: D_{R_{k}}^{-1}
\right\}\right]
}_{\mbox{\scriptsize Pre-array ${\mathbb D}_{A}$}}
\mathfrak{W}
= \!\!\! \underbrace{
\begin{bmatrix}
D^{-1}_{P_{k|k}}
\end{bmatrix}
}_{\mbox{\scriptsize Post-array ${\mathbb D}_{B}$}} \stackrel{\mbox{\scriptsize read-off}}{\Longrightarrow} \left[D_{P_{k|k}}^{-1}\right];$
\li \>Compute $K_{k}  = \lambda_k \left([\bar U_{P_{k|k}}^{-T}][D_{P_{k|k}}^{-1}][\bar U_{P_{k|k}}^{-T}]^T\right)^{-1}H_k^T R_k^{-1}$;
\li \>$\hat x_{k|k}  =   \hat x_{k|k-1}+K_{k}(y_k-H_k \hat x_{k|k-1})$.
\li \>Build pre-arrays ${\mathbb A}$, ${\mathbb D}_A$ and apply MWGS algorithm: \label{mcc:ud:f:P}
\zi \>$\underbrace{
\begin{bmatrix}
(I - K_{k}H_k)\bar U_{P_{k|k-1}} & \; K_k\bar U_{R_{k}}
\end{bmatrix}
}_{\mbox{\scriptsize Pre-array ${\mathbb A}^T$}}\!
 =\!\!\!
  \underbrace{
\begin{bmatrix}
\bar U_{P_{k|k}}
\end{bmatrix}
}_{\mbox{\scriptsize Post-array ${\mathbb B}$}}\!\!\!{\mathfrak Q}^T \stackrel{\mbox{\scriptsize read-off}}{\Longrightarrow} \left[\bar U_{P_{k|k}}\right];$
\zi \>$\mathfrak{Q}^T
\underbrace{
\left[{\rm diag}\left\{
D_{P_{k|k-1}}, \: D_{R_{k}}
\right\}\right]
}_{\mbox{\scriptsize Pre-array ${\mathbb D}_{A}$}}
\mathfrak{Q}
= \!\!\! \underbrace{
\begin{bmatrix}
D_{P_{k|k}}
\end{bmatrix}
}_{\mbox{\scriptsize Post-array ${\mathbb D}_{B}$}} \stackrel{\mbox{\scriptsize read-off}}{\Longrightarrow} \left[D_{P_{k|k}}\right].$
\end{codebox}

To validate the algorithm above, we start with the MWGS orthogonalization in line~\ref{mcc:ud:p:P} of Algorithm~1b, i.e.
\begin{align*}
&<[F_{k-1} \bar U_{P_{k-1|k-1}} \quad G_{k-1}\bar U_{Q_{k-1}}]><[F_{k-1} \bar U_{P_{k-1|k-1}} \quad G_{k-1}\bar U_{Q_{k-1}}]>_{{\mathbb D}_{A}} \\
& \qquad = X{\mathbb D}_{X}X^T \quad \mbox{ where } \quad {\mathbb D}_{A} = {\rm diag}\left\{ D_{P_{k-1|k-1}}, D_{Q_{k-1}}\right\},
\end{align*}
i.e. the following equation holds
\begin{align*}
X{\mathbb D}_{X}X^T & = F_{k-1} \underbrace{\bar U_{P_{k-1|k-1}}D_{P_{k-1|k-1}}\bar U_{P_{k-1|k-1}}^T}_{P_{k-1|k-1}}F_{k-1}^T \\
 & + G_{k-1}(\bar U_{Q_{k-1}}D_{Q_{k-1}}\bar U_{Q_{k-1}}^T) G_{k-1}^T = P_{k|k-1}
 \end{align*}
where the process covariance $Q_{k-1}$ is UD-factorized as well, i.e. $Q_{k-1} = \bar U_{Q_{k-1}}D_{Q_{k-1}}\bar U_{Q_{k-1}}^T$. Hence, we conclude $X:=\bar U_{P_{k|k-1}}$ and ${\mathbb D}_{X}:=D_{P_{k|k-1}}$.

At the same manner, the formulas in lines~\ref{mcc:ud:f:K} and~\ref{mcc:ud:f:P} of Algorithm~1b are validated, i.e.
\begin{align*}
&<[\bar U_{P_{k|k-1}}^{-T} \quad \lambda_k^{1/2}H_{k}^T\bar U_{R_{k}}^{-T}]><[\bar U_{P_{k|k-1}}^{-T} \quad \lambda_k^{1/2}H_{k}^T\bar U_{R_{k}}^{-T}]>_{{\mathbb D}_{A}} \\
& \qquad = Y{\mathbb D}_{Y}Y^T \quad \mbox{ where } \quad {\mathbb D}_{A} = {\rm diag}\left\{ D_{P_{k|k-1}}^{-1}, D_{R_{k}}^{-1}\right\}, \\
&<[(I - K_{k}H_k)\bar U_{P_{k|k-1}} \quad K_k\bar U_{R_{k}}],[(I - K_{k}H_k)\bar U_{P_{k|k-1}} \quad K_k\bar U_{R_{k}}]> \\
& \qquad = Z{\mathbb D}_{Z}Z^T \quad \mbox{ where } \quad {\mathbb D}_{A} = {\rm diag}\left\{ D_{P_{k|k-1}}, D_{R_{k}}\right\}
\end{align*}
and the measurement covariance matrix $R_{k}$ is UD-factorized, i.e. $R_{k} = \bar U_{R_k}D_{R_k}\bar U_{R_k}^T$. Thus, we get
\[
Y{\mathbb D}_{Y}Y^T  = \underbrace{\bar U_{P_{k|k-1}}^{-T} D_{P_{k|k-1}}^{-1} \bar U_{P_{k|k-1}}^{-1}}_{P_{k|k-1}^{-1}} + \lambda_k H_{k}^T
\underbrace{\bar U_{R_{k}}^{-T} D_{R_{k}}^{-1} \bar U_{R_{k}}^{-1}}_{R_{k}^{-1}}H_k.
\]
Having compared equation above with formula~\eqref{mcc:P:eq1}, we conclude that $Y:=\bar U_{P_{k|k}}^{-T}$ and ${\mathbb D}_{Y}:=D_{P_{k|k}}^{-1}$. Next,
\begin{align*}
Z{\mathbb D}_{Z}Z^T & = (I - K_{k}H_k)\!\underbrace{\bar U_{P_{k|k-1}} D_{P_{k|k-1}} \bar U_{P_{k|k-1}}^T}_{P_{k|k-1}}(I - K_{k}H_k)^T \\
& + K_{k}(\bar U_{R_{k}} D_{R_{k}} \bar U_{R_{k}}^{T})K_k^T = P_{k|k}
\end{align*}
and, hence, $Z:=\bar U_{P_{k|k}}$ and ${\mathbb D}_{Z}:=D_{P_{k|k}}$. This completes the proof of algebraic equivalence between the original MCC-KF (Algorithm~1) and its UD-based counterpart (Algorithm~1b).

Finally, we suggest the UD-based IMCC-KF implementation in Algorithm~2. In fact, it requires one less MWGS orthogonalization at each step because the re-calculation of $P_{k|k}$ via the Joseph stabilized equation is not required in the IMCC-KF.
\begin{codebox}
\Procname{{\bf Algorithm 2b}. $\proc{UD IMCC-KF}$ ({\it UD-based IMCC-KF})}
\zi \textsc{Initialization:}($k=0$) $\hat x_{0|0} = \bar x_0$ and $\bar U_{P_{0|0}} = \bar U_{\Pi_0}$, $D_{P_{0|0}} = D_{\Pi_0}$
\zi \>where UD-decomposition is applied: $\Pi_0 = \bar U_{\Pi_0}D_{\Pi_0}\bar U_{\Pi_0}^T$.
\zi \textsc{Time Update}: ($k=\overline{1,N}$) Repeat lines~1,2 of Algorithm~1b;
\zi \textsc{Measurement Update}: ($k=\overline{1,N}$)
\li Compute $\lambda_k$ by formula~\eqref{eq:lambda};
\li Build pre-arrays ${\mathbb A}$, ${\mathbb D}_A$ and apply MWGS algorithm: \label{imcc:ud:f:P}
\zi $\underbrace{
\begin{bmatrix}
\bar U_{P_{k|k-1}} & 0\\
\lambda_k^{1/2}H_k\bar U_{P_{k|k-1}} & \bar U_{R_k}
\end{bmatrix}
}_{\mbox{\scriptsize Pre-array ${\mathbb A}^T$}}
\! =\!
  \underbrace{
\begin{bmatrix}
\bar U_{P_{k|k}} & \bar K_{k}^u\\
0 & \bar U_{R_{e,k}}
\end{bmatrix}
}_{\mbox{\scriptsize Post-array ${\mathbb B}$}}\!{\mathfrak Q}^T \!\!\!\!
\begin{matrix}
\!\!\!\!\!\stackrel{\mbox{\scriptsize read-off}}{\Longrightarrow} \left[\bar U_{P_{k|k}}\right]; \\
\quad \Longrightarrow [\bar U_{R_{e,k}}], [\bar K_{k}^u];
\end{matrix}$
\zi $\mathfrak{Q}^T\!
\underbrace{
\left[{\rm diag}\left\{
D_{P_{k|k-1}}, \: D_{R_{k}}
\right\}\right]
}_{\mbox{\scriptsize Pre-array ${\mathbb D}_{A}$}}\!\!
\mathfrak{Q}
= \! \underbrace{
{\rm diag}\left\{
D_{P_{k|k}}, \: D_{R_{e,k}}
\right\}
}_{\mbox{\scriptsize Post-array ${\mathbb D}_{B}$}} \!\!\stackrel{\mbox{\scriptsize read-off}}{\Longrightarrow}\!\! \left[D_{P_{k|k}}\right];$
\li $\hat x_{k|k}  =   \hat x_{k|k-1}+\lambda_k^{1/2}\left[\bar K_{k}^u\right]\left[\bar U_{R_{e,k}}\right]^{-1}(y_{k}-H\hat x_{k|k-1})$. \label{imcc:ud:f:X}
\end{codebox}

The method in Algorithm~2b can be validated at the same manner as Algorithm~1b. More precisely, the time update step of these implementations is the same and, hence, it has been already proved. Let's consider the transformation in line~\ref{imcc:ud:f:P}  of Algorithm~2b. We have the following set of equations
\begin{align*}
&<[\bar U_{P_{k|k-1}} \quad  0]><[\bar U_{P_{k|k-1}} \quad  0]>_{{\rm diag}\left\{
D_{P_{k|k-1}}, \: D_{R_{k}}\right\}} \\
& \quad = <\![X \quad \!\!Y],[X \quad \!\! Y]\!>_{{\rm diag}\left\{
D_1, \: D_2\right\}}, \\
&<[\bar U_{P_{k|k-1}} \quad  0]><[\lambda_k^{1/2}H_k\bar U_{P_{k|k-1}} \quad \bar U_{R_k}]>_{{\rm diag}\left\{
D_{P_{k|k-1}}, \: D_{R_{k}}\right\}} \\
& \quad = <\![X \quad \!\!Y],[0 \quad \!\! Z]\!>_{{\rm diag}\left\{
D_1, \: D_2\right\}}, \\
&<\![\lambda_k^{1/2}H_k\bar U_{P_{k|k-1}} \quad \bar U_{R_k}],[\lambda_k^{1/2}H_k\bar U_{P_{k|k-1}} \quad \bar U_{R_k}]\!>_{{\rm diag}\left\{
D_{P_{k|k-1}}, \: D_{R_{k}}\right\}} \\
&\quad = <\![0 \quad \!\!Z],[0 \quad \!\! Z]\!>_{{\rm diag}\left\{
D_1, \: D_2\right\}}\!.
\end{align*}

From the last equation, we have
\[
ZD_2Z^T  = \lambda_kH_{k}\underbrace{(\bar U_{P_{k|k-1}} D_{P_{k|k-1}} \bar U_{P_{k|k-1}}^T)}_{P_{k|k-1}}H_{k}^T+\underbrace{(\bar U_{R_k} D_{R_k} \bar U_{R_k}^T)}_{R_{k}} = R_{e,k},
\]
i.e. $Z:=\bar U_{R_{e,k}}$ and $D_2 := D_{R_{e,k}}$.

Having substituted the resulted $Z$ and $D_2$ values into the second equation in the set above, we obtain
\[ YD_{R_{e,k}}\bar U_{R_{e,k}}^T = \lambda_k^{1/2}(\bar U_{P_{k|k-1}} D_{P_{k|k-1}} \bar U_{P_{k|k-1}}^T)H_{k}^T.
\]
Hence, $Y := \lambda_k^{1/2} P_{k|k-1}H_k^T\bar U_{R_{e,k}}^{-T}D_{R_{e,k}}^{-1}$. This value is denoted as $\bar K_k^{u}$ in Algorithm~2b, i.e. $Y:=\bar K_k^{u}$. From formula~\eqref{mcc:K:eq1}, the following relationship is obtained: $K_k = \lambda_k P_{k|k-1}H_k^TR_{e,k}^{-1} = \lambda_k^{1/2}\bar K_k^{u} \bar U_{R_{e,k}}^{-1}$. Next, we note that the block $\bar K_k^{u}$ is directly read-off from post-array in Algorithm~2b. Hence, it makes sense to use it for computing state estimate, straightforward
\begin{align*}
\hat x_{k|k} & =   \hat x_{k|k-1}+K_{k}(y_k-H_k \hat x_{k|k-1}) \\
&=  \hat x_{k|k-1}+\lambda_k^{1/2}\bar K_k^{u} \bar U_{R_{e,k}}^{-1}(y_k-H_k \hat x_{k|k-1}),
\end{align*} i.e. the formula in line~\ref{imcc:ud:f:X} of Algorithm~2b is validated.

Finally, we consider the last relationship that is
 \[XD_1X^T + \bar K_k^{u} D_{R_{e,k}} (\bar K_k^{u})^T = P_{k|k-1},
  \]
  i.e. we have
\begin{align*}
XD_1X^T & = P_{k|k-1}-\lambda_k^{1/2} P_{k|k-1}H_k^T\bar U_{R_{e,k}}^{-T}D_{R_{e,k}}^{-1}D_{R_{e,k}}(\lambda_k^{-1/2}K_k\bar U_{R_{e,k}})^T \\
& = P_{k|k-1} - P_{k|k-1}H_k^TK_k^T = P_{k|k-1}(I - H_k^TK_k^T) = P_{k|k}^T.
\end{align*}
Taking into account a symmetric form of any covariance matrix, we conclude $X:=\bar U_{P_{k|k}}$ and $D_1:= D_{P_{k|k}}$. This completes the proof of the UD-based IMCC-KF (Algorithm~2b).

\section{SVD-based factored-form implementations}  \label{UD:filters}

To the best of author's knowledge, there exist only two {\it classical} KF methods based on the singular value decomposition (SVD). The first SVD-based KF was developed in~\cite{WangSVD1992}. However, it was shown to be numerically unstable with respect to roundoff errors. Thereby, the robust SVD-based KF algorithm has been recently proposed in~\cite{2017:IET:KulikovaTsyganova}. In this paper, the goal is to extend  the SVD-based filtering on the MCC-KF (Algorithm~1) and IMCC-KF (Algorithm~2) techniques.

Each iteration of the new filtering methods is implemented by using the SVD factorization~\cite[Theorem~1.1.6]{Bjorck2015}: Every matrix $A \in {\mathbb C}^{m\times n}$ of rank $r$ can be written as follows:
\[
A  = W\Sigma V^*, \,
\Sigma =
\begin{bmatrix}
S & 0 \\
0 & 0
\end{bmatrix} \in {\mathbb R}^{m\times n},  \; S={\rm diag}\{ \sigma_1,\ldots,\sigma_r\}
\]
where $W \in {\mathbb C}^{m\times m}$, $V \in {\mathbb C}^{n\times n}$  are unitary matrices, $V^*$ is the conjugate transpose of $V$, and $S \in {\mathbb R}^{r\times r}$ is a real nonnegative diagonal matrix. Here $\sigma_1\geq \sigma_2\geq \ldots \geq\sigma_r>0$ are called the
singular values of $A$. (Note that if $r = n$ and/or $r = m$, some of the zero submatrices in $\Sigma$ are empty.)

The SVD-based filtering methods belong to the {\it factored-form} (square-root) family, because the covariance $P$ is factorized in the form $P = V D V^T$  where $V$ is an orthogonal matrix and $D$ is a diagonal matrix with singular values of $P$. Hence, one can set the matrix square root as follows: $P = SS^T$ where $S:=V D^{1/2}$. It is also worth noting here that in the SVD-based filtering, the square-root factor $S$ is a full matrix, in general. This is in contrast to the upper or lower triangular factor $S$ in the Cholesky-based implementations discussed in previous sections. Additionally, the SVD factorization provides the users with extra information about the matrix structure and properties and, hence, it might be used in various reduced-rank filtering design strategies.

The following SVD-based variant for the MCC-KF estimator (Algorithm~1) is proposed.
\begin{codebox}
\Procname{{\bf Algorithm 1c}. $\proc{SVD MCC-KF}$ ({\it SVD-based MCC-KF})}
\zi \textsc{Initialization:}($k=0$) $\hat x_{0|0} = \bar x_0$ and $V_{P_{0|0}} = V_{\Pi_0}$, $D^{1/2}_{P_{0|0}} = D^{1/2}_{\Pi_0}$
\zi \>where SVD-decomposition is applied: $\Pi_0 = V_{\Pi_0}D_{\Pi_0}V_{\Pi_0}^T$.
\zi \textsc{Time Update}: ($k=\overline{1,N}$)
\li \>$\hat x_{k|k-1}  = F_{k-1} \hat x_{k-1|k-1}$;
\li \>Build pre-array ${\mathbb A}$ and apply SVD factorization \label{mcc:svd:p:P}
\zi \>$\underbrace{
\begin{bmatrix}
D^{1/2}_{P_{k-1|k-1}}V^T_{P_{k-1|k-1}}F_{k-1}^T\\
D^{1/2}_{Q_{k-1}}V^T_{Q_{k-1}}G_{k-1}^T
\end{bmatrix}
}_{\rm Pre-array \: {\mathbb A}} \!\!
=\!
\underbrace{
\mathfrak{W}\!
\begin{bmatrix}
D_{P_{k|k-1}}^{1/2} \\
0
\end{bmatrix}\!\!
\mathfrak{V}^T
}_{\rm Post-arrays}
\begin{matrix}
\!\!\!\!\!\!\!\!\stackrel{\mbox{\scriptsize read-off}}{\Longrightarrow} \left[D_{P_{k|k-1}}^{1/2}\right]; \\
\Longrightarrow [V_{P_{k|k-1}}=\mathfrak{V}];
\end{matrix}$
\zi \textsc{Measurement Update}: ($k=\overline{1,N}$)
\li \>Compute $\lambda_k$ by formula~\eqref{eq:lambda};
\li \>Build pre-array ${\mathbb A}$ and apply SVD factorization \label{mcc:svd:f:K}
\zi \>$\underbrace{
\begin{bmatrix}
\lambda_k^{1/2}D^{-1/2}_{R_{k}}V_{R_{k}}^T H_kV_{P_{k|k-1}} \\
D_{P_{k|k-1}}^{-1/2}
\end{bmatrix}
}_{\rm Pre-array  \: {\mathbb A}}
\! \! =\!
\underbrace{
\mathfrak{W}\!
\begin{bmatrix}
D_{P_{k|k}}^{-1/2} \\
0
\end{bmatrix}\!
\mathfrak{V}^T
}_{\rm Post-arrays}
\!\!\!\!\!\!
\begin{matrix}
\quad \stackrel{\mbox{\scriptsize read-off}}{\Longrightarrow} \left[D_{P_{k|k}}^{-1/2}\right]; \\
\!\!\Longrightarrow [\mathfrak{V}];\end{matrix}
$
\li \>Calculate the SVD factor $V_{P_{k|k}}=V_{P_{k|k-1}}\mathfrak{V}$; \label{mcc:svd:f:Pinv}
\li \>Compute $K_{k}  = \lambda_k [V_{P_{k|k}}][D_{P_{k|k}}^{-1/2}]^{-2}[V_{P_{k|k}}]^{T}H_k^T R_k^{-1}$; \label{mcc:svd:f:K1}
\li \>$\hat x_{k|k}  =   \hat x_{k|k-1}+K_{k}(y_k-H_k \hat x_{k|k-1})$.
\li \>Build pre-array ${\mathbb A}$ and apply SVD factorization  \label{mcc:svd:f:P}
\zi \>$\underbrace{
\begin{bmatrix}
D_{P_{k|k-1}}^{1/2}V_{P_{k|k-1}}^T(I - K_k H_k)^T\\
D^{1/2}_{R_{k}}V^T_{R_{k}}K_{k}^T
\end{bmatrix}
}_{\rm Pre-array \: {\mathbb A}} \!\!
=\!
\underbrace{
\mathfrak{W}\!
\begin{bmatrix}
D_{P_{k|k}}^{1/2} \\
0
\end{bmatrix}\!\!
\mathfrak{V}^T
}_{\rm Post-arrays}
\begin{matrix}
\!\!\!\!\!\!\!\!\stackrel{\mbox{\scriptsize read-off}}{\Longrightarrow} \left[D_{P_{k|k}}^{1/2}\right]; \\
\Longrightarrow [V_{P_{k|k}}=\mathfrak{V}].
\end{matrix}$
\end{codebox}

\begin{rem}
In the algorithm above, the SVD is applied to the process and measurement noise covariances as well, i.e $Q_k = V_{Q_k}D_{Q_k}V_{Q_k}^T$ and $R_k = V_{R_k}D_{R_k}V_{R_k}^T$. However, as discussed in~\cite[Remark~2]{2017:IET:KulikovaTsyganova}, the Cholesky decomposition might be used for $Q_k$ and $R_k$ instead of the SVD factorization.
\end{rem}

To prove the algebraic equivalence between the new SVD-based implementation (Algorithm~1c) and the original MCC-KF (Algorithm~1), we note that $A^TA = (V \Sigma W^T)(W\Sigma V^T)^T = V \Sigma^2 V^T$. Having compared both sides of the resulted equality $A^TA=V \Sigma^2 V^T$, one may validate the formulas in Algorithm~1c. Indeed, from the factorization in line~\ref{mcc:svd:p:P}, we obtain
\begin{align*}
& F_{k-1}\underbrace{V_{P_{k-1|k-1}}D_{P_{k-1|k-1}}V_{P_{k-1|k-1}}^T}_{P_{k-1|k-1}} F_{k-1}^T + G_{k-1}\underbrace{V_{Q_{k-1}}D_{Q_{k-1}} V^T_{Q_{k-1}}}_{Q_{k-1}}G_{k-1}^T \\
& = \mathfrak{V} D_{P_{k|k-1}}\mathfrak{V} ^T = P_{k|k-1}, \mbox{ i.e. } \mathfrak{V} := V_{P_{k|k-1}}.
\end{align*}

Next, in line~\ref{mcc:svd:f:K} of Algorithm~1c we have
\begin{equation}
\label{prove:1}
\lambda_k V_{P_{k|k-1}}^TH_k^T\underbrace{V_{R_{k}} D^{-1}_{R_{k}}V_{R_{k}}^T}_{R_k^{-1}}H_kV_{P_{k|k-1}} + D_{P_{k|k-1}}^{-1} = \mathfrak{V} D^{-1}_{P_{k|k}}\mathfrak{V}^T.
\end{equation}

Having multiplied both sides of equation~\eqref{prove:1} by $V_{P_{k|k-1}}$ (at the left) and by $V_{P_{k|k-1}}^T$ (at the right), we get
\begin{equation}
\label{prove:2}
\lambda_k H_k^TR_k^{-1}H_k + \underbrace{V_{P_{k|k-1}}D_{P_{k|k-1}}^{-1}V_{P_{k|k-1}}^T}_{P_{k|k-1}^{-1}} = V_{P_{k|k-1}}\mathfrak{V} D^{-1}_{P_{k|k}}\mathfrak{V}^TV_{P_{k|k-1}}^T.
\end{equation}
Having compared formula~\eqref{prove:2} with equation~\eqref{mcc:P:eq1}, i.e. $P_{k|k} = \left( P_{k|k-1}^{-1}+\lambda_kH_k^TR_k^{-1}H_k\right)^{-1}$, we conclude that $V_{P_{k|k}}= V_{P_{k|k-1}}\mathfrak{V}$. This validates formula in line~\ref{mcc:svd:f:Pinv} of Algorithm~1c.

Next, formula for computing the gain $K_k$ in line~\ref{mcc:svd:f:K1} is the same as in the original MCC-KF (Algorithm~1) where the SVD is used for matrix $P_{k|k}$. Finally, factorization in line~\ref{mcc:svd:f:P} of Algorithm~1c implies the symmetric Joseph stabilized equation of the classical KF utilized in the MCC-KF (Algorithm~1), i.e.
\begin{align*}
A^TA & = K_kR_kK_k^T + \left(I - K_{k} H_k\right)\underbrace{V_{P_{k|k-1}}D_{P_{k|k-1}}V^T_{P_{k|k-1}}}_{P_{k|k-1}} \left(I - K_{k}H_k\right)^T \\
& = \mathfrak{V}D_{P_{k|k}}\mathfrak{V}^T = P_{k|k}, \mbox{ i.e. } \mathfrak{V} := V_{P_{k|k}}.
\end{align*}
This concludes the proof of Algorithm~1c.

Similar, the SVD-based IMCC-KF is derived and summarized in Algorithm~2.
\begin{codebox}
\Procname{{\bf Algorithm 2c}. $\proc{SVD IMCC-KF}$ ({\it SVD-based IMCC-KF})}
\zi \textsc{Initialization:}($k=0$) $\hat x_{0|0} = \bar x_0$ and $V_{P_{0|0}} = V_{\Pi_0}$, $D^{1/2}_{P_{0|0}} = D^{1/2}_{\Pi_0}$
\zi \>where SVD-decomposition is applied: $\Pi_0 = V_{\Pi_0}D_{\Pi_0}V_{\Pi_0}^T$.
\zi \textsc{Time Update}: ($k=\overline{1,N}$) Repeat lines~1,2 of Algorithm~1c;
\zi \textsc{Measurement Update}: ($k=\overline{1,N}$)
\li \>Compute $\lambda_k$ by formula~\eqref{eq:lambda};
\li \>Build pre-array ${\mathbb A}$ and apply SVD factorization \label{imcc:svd:f:P}
\zi \>$\underbrace{
\begin{bmatrix}
\lambda_k^{1/2}D^{-1/2}_{R_{k}}V_{R_{k}}^T H_kV_{P_{k|k-1}} \\
D_{P_{k|k-1}}^{-1/2}
\end{bmatrix}
}_{\rm Pre-array  \: {\mathbb A}}
\! \! =\!
\underbrace{
\mathfrak{W}\!
\begin{bmatrix}
D_{P_{k|k}}^{-1/2} \\
0
\end{bmatrix}\!
\mathfrak{V}^T
}_{\rm Post-arrays}
\!\!\!\!\!\!
\begin{matrix}
\quad \stackrel{\mbox{\scriptsize read-off}}{\Longrightarrow} \left[D_{P_{k|k}}^{-1/2}\right]; \\
\!\!\Longrightarrow [\mathfrak{V}];
\end{matrix}$
\li \>Calculate the SVD factor $V_{P_{k|k}}=V_{P_{k|k-1}}\mathfrak{V}$; \label{imcc:svd:f:Pinv}
\li \>Compute $K_{k}  = \lambda_k [V_{P_{k|k}}][D_{P_{k|k}}^{-1/2}]^{-2}[V_{P_{k|k}}]^{T}H_k^T R_k^{-1}$; \label{imcc:svd:f:K}
\li \>$\hat x_{k|k}  =   \hat x_{k|k-1}+K_{k}(y_k-H_k \hat x_{k|k-1})$.
\end{codebox}

As in all previous methods, the time update steps in Algorithms~1c and~2c coincide. Next, the IMCC-KF requires one less SVD factorization at each filtering step because it does not demand the re-computation of $P_{k|k}$ by the symmetric Joseph stabilized equation at the end of each iterate. Thus, the IMCC-KF implementation (Algorithm~2c) is, in fact, the MCC-KF Algorithm~1c without the last $P_{k|k}$ re-calculation, i.e. all formulas of Algorithm~2c have been already proved in this section.

Our final remark concerns the robust variant of the SVD-based implementations. Both Algorithm~1c and Algorithm~2c require two matrix inversions that are related to $D_{P_{k|k-1}}^{-1}$ and $D_{P_{k|k}}^{-1}$ calculation. For numerical stability and computational complexity reasons, it is preferable to avoid this operation. Following~\cite{2017:IET:KulikovaTsyganova}, we can suggest  the robust SVD-based implementation for the MCC-KF as follows.
\begin{codebox}
\Procname{{\bf Algorithm 1d}. $\proc{rSVD MCC-KF}$ ({\it robust SVD-based MCC-KF})}
\zi \textsc{Initialization:}($k=0$) $\hat x_{0|0} = \bar x_0$ and $V_{P_{0|0}} = V_{\Pi_0}$, $D^{1/2}_{P_{0|0}} = D^{1/2}_{\Pi_0}$
\zi \>where SVD-decomposition is applied: $\Pi_0 = V_{\Pi_0}D_{\Pi_0}V_{\Pi_0}^T$.
\zi \textsc{Time Update}: ($k=\overline{1,N}$) Repeat lines~1,2 of Algorithm~1c;
\zi \textsc{Measurement Update}: ($k=\overline{1,N}$)
\li \>Compute $\lambda_k$ by formula~\eqref{eq:lambda};
\li \>Build pre-array ${\mathbb A}$ and apply SVD factorization \label{mcc:svd:f:Rek}
\zi \>$\underbrace{
\begin{bmatrix}
\lambda_k^{1/2} D^{1/2}_{P_{k|k-1}} V_{P_{k|k-1}}^TH_k^T \\
D^{1/2}_{R_k}V_{R_k}^T
\end{bmatrix}
}_{\rm Pre-array  \: {\mathbb A}}
\! \! =\!
\underbrace{
\mathfrak{W}\!
\begin{bmatrix}
D_{R_{e,k}}^{1/2} \\
0
\end{bmatrix}\!
\mathfrak{V}^T
}_{\rm Post-arrays}
\!\!\!\!\!\!
\begin{matrix}
\quad \stackrel{\mbox{\scriptsize read-off}}{\Longrightarrow} \left[D_{R_{e,k}}^{1/2}\right]; \\
\!\!\Longrightarrow [V_{R_{e,k}}=\mathfrak{V}];
\end{matrix}$
\li \>Compute $K_{k}  = \lambda_k P_{k|k-1}H_k^T [V_{R_{e,k}}] [D_{R_{e,k}}^{1/2}]^{-2} [V_{R_{e,k}}]^T$; \label{mcc:svd:f:K:d}
\li \>$\hat x_{k|k}  =   \hat x_{k|k-1}+K_{k}(y_k-H_k \hat x_{k|k-1})$.
\li \>Build pre-array ${\mathbb A}$ and apply SVD factorization  \label{mcc:svd:f:P:d}
\zi \>$\underbrace{
\begin{bmatrix}
D_{P_{k|k-1}}^{1/2}V_{P_{k|k-1}}^T(I - K_k H_k)^T\\
D^{1/2}_{R_{k}}V^T_{R_{k}}K_{k}^T
\end{bmatrix}
}_{\rm Pre-array \: {\mathbb A}} \!\!
=\!
\underbrace{
\mathfrak{W}\!
\begin{bmatrix}
D_{P_{k|k}}^{1/2} \\
0
\end{bmatrix}\!\!
\mathfrak{V}^T
}_{\rm Post-arrays}
\begin{matrix}
\!\!\!\!\!\!\!\!\stackrel{\mbox{\scriptsize read-off}}{\Longrightarrow} \left[D_{P_{k|k}}^{1/2}\right]; \\
\Longrightarrow [V_{P_{k|k}}=\mathfrak{V}].
\end{matrix}$
\end{codebox}

The difference between Algorithms~1c and~1d is in the gain matrix $K_k$ calculation. Algorithm~1c utilizes equation~\eqref{mcc:K:eq2} while Algorithm~1d implies formula~\eqref{mcc:K:eq1}, i.e. $K_k = \lambda_k P_{k|k-1}H_k^T\left(\lambda_kH_kP_{k|k-1}H_k^T+R_k\right)^{-1}$. The matrix that needs to be inverted is denoted as $R_{e,k}$, i.e. $R_{e,k} = \lambda_kH_kP_{k|k-1}H_k^T+R_k$. Its SVD factors are computed in line~\ref{mcc:svd:f:Rek} of Algorithm~1d, i.e.
\[
A^TA = \lambda_kH_k \underbrace{V_{P_{k|k-1}}D_{P_{k|k-1}}V_{P_{k|k-1}}^T}_{P_{k|k-1}} H_k^T+\underbrace{V_{R_k}D_{R_k}V_{R_k}^T}_{R_k} = V_{R_{e,k}}D_{R_{e,k}}V^T_{R_{e,k}}.
\]

When factors $V_{R_{e,k}}$ and $D_{R_{e,k}}$ are computed, the gain matrix can be found as follows:
\[
K_k = \lambda_k P_{k|k-1}H_k^TR_{e,k}^{-1} = \lambda_k P_{k|k-1}H_k^T [V_{R_{e,k}}] [D_{R_{e,k}}^{1/2}]^{-2} [V_{R_{e,k}}]^T.
\]
This validates the computation in line~\ref{mcc:svd:f:K:d} of Algorithm~1d. Finally, the underlying formula for the $P_{k|k}$ calculation is  the symmetric Joseph stabilized equation, i.e. it is the same as in Algorithm~1c. This completes the proof of Algorithm~1d.

As discussed in~\cite{2017:IET:KulikovaTsyganova}, such organization of computations (see Algorithm~1d) improves numerical stability of SVD-based filtering with respect to roundoff errors, because less matrix inversions are required. More precisely, only $D_{R_{e,k}}^{-1}$ is involved in Algorithm~1d, while  $D_{P_{k|k-1}}^{-1}$ and $D_{P_{k|k}}^{-1}$ are not required.

\begin{table*}
\caption{Theoretical comparison of the factored-form implementations developed for the MCC-KF (Algorithm~1) and IMCC-KF (Algorithm~2) estimators.} \label{tab:properties}
\centering
{\small
\begin{tabular}{l||c|c||c|c||c|c|c}
\hline
{\bf Property} &
\multicolumn{2}{c||}{\bf Cholesky-based methods} & \multicolumn{2}{c||}{\bf UD-based methods} & \multicolumn{3}{c}{\bf SVD-based methods}\\
\cline{2-8}
  &  MCC-KF & IMCC-KF &  MCC-KF & IMCC-KF &  \multicolumn{2}{c|}{MCC-KF} &  IMCC-KF \\
\hhline{*{5}{|~}*{2}{|-}|~|}
  &  Alg.~1a & \cite[Alg.~2]{2017:CSL:Kulikova} &  Alg.~1b & Alg.~2b &  Alg.~1c &  Alg.~1d & Alg.~2c \\
\hline
1. Type          & Covariance     & Covariance     & Covariance & Covariance & Covariance & Covariance & Covariance\\
2. Decomposition & Cholesky       & Cholesky       & \multicolumn{2}{c||}{modified Cholesky}  & SVD & SVD  & SVD \\
3. Restriction   & \multicolumn{2}{c||}{$\Pi_0>0$, \quad $Q_k>0$, \quad $R_k>0$} &
                          \multicolumn{2}{c||}{$\Pi_0>0$, \quad $Q_k>0$, \quad $R_k>0$} & no & no  & no  \\
4. Pre-array               & TU: any QR(1) & any QR(1) & MWGS(1)  & MWGS(1)& SVD (1) & SVD (1) & SVD (1) \\
\phantom{4.} factorization & MU: any QR(2) & any QR(1) & MWGS(2)  & MWGS(1)& SVD (2) & SVD (2) & SVD (1) \\
5. Matrix        & $P_{k|k-1}^{-1/2}$, $P_{k|k}^{-1/2}$       & $R_{e,k}^{-1/2}$ & $\bar U_{P_{k|k-1}}^{-1}$, $\bar U_{P_{k|k}}^{-1}$, &
$\bar U_{R_{e,k}}^{-1}$ &  $D_{P_{k|k-1}}^{-1/2}$, $D_{P_{k|k}}^{-1/2}$ &  $D_{R_{e,k}}^{-1/2}$ & $D_{P_{k|k-1}}^{-1/2}$, $D_{P_{k|k}}^{-1/2}$  \\
\phantom{5.} inversions    &    &                                       &                   $D_{P_{k|k-1}}^{-1}$, $D_{P_{k|k}}^{-1}$  &          &
                           &                                                                    &  \\
6. Extended form & -- & \cite[Alg.~3]{2017:CSL:Kulikova} & --  & ?   & --  &   --  & ? \\
  \hline
\end{tabular}
}
\end{table*}

Unfortunately, it is not possible to design the similar robust SVD-based variant for the IMCC-KF (Algorithm~2). More precisely, the goal is to avoid $D_{P_{k|k-1}}^{-1}$ and $D_{P_{k|k}}^{-1}$ operations in Algorithm~2c. Clearly, the gain computation $K_k$ can be performed at the same way as it is done in Algorithm~1d, i.e. it requires $D_{R_{e,k}}^{-1}$ computation, additionally. Having computed $K_k$, the SVD factors of the error covariance matrix $P_{k|k}$ should be updated through SVD factorization of the related pre-array, say $A$. For that, the underlying equation for $P_{k|k}$ should have a symmetric form, because any covariance matrix is symmetric, i.e. the related SVD is apllied to the symmetric pre-arrays product $A^TA$ or $AA^T$. For the IMCC-KF method, there are three possibilities for computing $P_{k|k}$, given by equations~\eqref{mcc:P:eq1}~-- \eqref{mcc:P:eq3}. Only formula~\eqref{mcc:P:eq1} has the required symmetric form and it is already used in Algorithm~2c implying the calculation of $D_{P_{k|k-1}}^{-1}$ and $D_{P_{k|k}}^{-1}$.  Equation~\eqref{mcc:P:eq3} might be symmetric if one skips the scalar parameter $\lambda_k$. However, in this case the Joseph stabilized equation is obtained, i.e. we get the MCC-KF implementation (Algorithm~1 and its SVD-based variant in Algorithm~1d), but not the IMCC-KF method in Algorithm~2.  The author still does not know how to balance equation~\eqref{mcc:P:eq3}, i.e. to express it in symmetric form that is appropriate for deriving the robust SVD-based implementation of the IMCC-KF (Algorithm~2). This is an open question for a future research.

\section{Discussion and comparison}

\subsection{Theoretical comparison}

Table~\ref{tab:properties} illustrates some theoretical aspects of the suggested factored-form filters' families. The following notation is used: sign ``+'' means that the corresponding property is available, sign ``--'' implies  missing corresponding feature, and ``?'' means that
the factored-form implementation with the related property might be derived in future.

Having analyzed the information presented in Table~\ref{tab:properties}, we make the following conclusions. First, all filtering algorithms developed in this paper are of {\it covariance}-type. This means that the error covariance matrix $P_{k|k}$ (or its factors) are updated according to the underlying filter recursion. An alternative class of methods implies $\Lambda_{k|k}=P_{k|k}^{-1}$ propagation (called the information matrix) rather than $P_{k|k}$. Such algorithms are known as {\it information}-type implementations and they have some benefits over the covariance recursions. One of the main reason to derive such implementations is a need to solve the state estimation problem without a {\it prior} information. In this case the initial error covariance matrix $\Pi_0$ is too ``large'' and, hence, the initialization step $\Pi_0 := \infty$ yields various complications for covariance filtering, while the information algorithms simply imply $\Lambda_0 := 0$. Additionally, the information filtering suggests a possible solution to numerical instability problem caused by influence of roundoff errors at the measurement update stage as discussed in~\cite[p.~356-357]{GrewalAndrews2015}. Further argument for deriving information filter recursion under the MCC approach is the matrix inversion $P_{k|k-1}^{-1}$ required at each iterate while computing the inflation parameter $\lambda_k$. To avoid this operation, it might be useful to derive the algebraic equivalent counterpart that updates the inverse (information) matrices (or their factors) automatically. It is worth noting here that similar motivation was used for developing information filtering for the classical KF in~\cite{Dyer1969}. All these facts make the information-type implementations attractive for practical use. To the best of the author's knowledge, the information MCC KF-like methods still do not exist. Their derivation could be an area for a future research.

The second row in Table~\ref{tab:properties} summarizes the decomposition of covariance matrices involved in each implementation under examination. The type of factorization may impose restrictions on their properties. For instance, the Cholesky decomposition is known to exist and to be unique when the symmetric matrix to be decomposed is positive definite~\cite{Golub1983}. These conditions are presented in the third row of Table~\ref{tab:properties}. In general,  covariance is a positive semi-definite matrix and the Cholesky decomposition still exists for such matrices, however, it is not unique~\cite{Higham1990}. In this case, the Cholesky-based implementations are unexpectedly interrupted by the procedure performing the decomposition. From this point of view, the SVD-based filtering might be preferable because no restrictions are implied for performing SVD; see~\cite[Theorem~1.1.6]{Bjorck2015}. Additionally, the SVD is the most accurate method to factorize the error covariance matrix (especially when it is close to singular), although it is more time consuming than the Cholesky decomposition.

Concerning the computational time, we conclude that the factored-form IMCC-KF implementations (Algorithms~2b, 2c and the previously published Algorithm~2 in~\cite{2017:CSL:Kulikova}) are expected to work faster than the factored-form MCC-KF counterparts (Algorithms~1a, 1b, 1c, 1d), because they require less QR/MWGS/SVD factorizations at each filtering step. The precise number of computations required by each implementation depends on a particular QR, square-root-free QR (QRD algorithms, e.g.~\cite{1991:Gotze,1993:Hsieh}) and SVD methods utilized in practice. While deriving the factored-form implementations, no restriction on the pre-array transformations is imposed, i.e. the rotations can be implemented in various ways and, hence, the
computational complexity analysis heavily depends on the users' choice and QR/MWGS/SVD method implemented.


Next, the required matrix inversions are outlined for each filtering algorithm in Table~1. As it has been already mentioned, it is preferable to avoid this operation in practice, for numerical and computational complexity reasons. The matrix $R_k^{-1}$ required in calculating $\lambda_k$ are not presented in Table~\ref{tab:properties}, because this part is the same for all implementations, i.e. we take it out of the consideration. Having analyzed the information presented in Table~\ref{tab:properties}, we conclude that the Cholesky- and UD-based IMCC-KF (see~\cite[Algorithm~2]{2017:CSL:Kulikova} and Algorithm~2b) are expected to possess a better numerical behavior than their MCC-KF counterparts (Algorithms~1a and~1b), because they require the inverse of $R_{e,k} \in {\mathbb R}^{m\times m}$ factors, only. Meanwhile Algorithms~1a and~1b involve the inverse of the error covariances $P_{k|k-1}$, $P_{k|k}  \in {\mathbb R}^{n\times n}$  factors. Besides, if $n>>m$, then the Cholesky- and UD-based IMCC-KF algorithms are expected to be faster than the MCC-KF analogues (Algorithms~1a and 1b). However, for the SVD-based implementations this is not the case. Indeed, both Algorithm~1c and~2c are expected to be of the same robustness with respect to roundoff errors, because they imply the scalar divisions by square roots of the same singular values; see the terms $D_{P_{k|k-1}}^{-1/2}$, $D_{P_{k|k}}^{-1/2}$ in Algorithms~1c and~2c. In contrast, Algorithm~1d demands the inversion of diagonal matrix $D^{1/2}_{R_{e,k}}$, only. Thus, it is expected to be more numerically stable with respect to roundoff errors than other SVD-based implementations, i.e. Algorithms~1c and~2c. In summary, only one SVD-based implementation has been found for the IMCC-KF estimator (Algorithm~2c). Meanwhile, two SVD-based implementations   have been developed for the MCC-KF estimator (Algorithms~1c and 1d). Among these two methods, one implementation (Algorithm~1d) is expected to be the most numerically robust with respect to roundoff errors.

The final remark concerns the state vector $\hat x_{k|k}$ computation. We stress that the so-called {\it extended} array form is practically feasible for Cholesky-based IMCC-KF and it has been recently published in~\cite[Algorithm~3]{2017:CSL:Kulikova}. The key idea of such methods comes from the KF community where the extended array implementations exist for the Cholesky-based filtering~\cite{1995:ParkKailath} and for the UD-based methods~\cite{1991:Chun,2017:Semushin} while for the SVD-based algorithms this problem is still open~\cite{2017:IET:KulikovaTsyganova}. The extended form implies an orthogonal transformation of {\it augmented} pre-array $[{\mathbb A}\: | \: b]$. As a result, instead of explicit formula $\hat x_{k|k}  =   \hat x_{k|k-1}+K_{k}(y_k-H_k \hat x_{k|k-1})$ for computing the state estimate, one utilizes a simple multiplication $\hat x_{k|k}  = \left[P_{k|k}^{T/2}\right] \left[P^{-T/2}_{k|k}\hat x_{k|k}\right]$ of the blocks $\left[P_{k|k}^{1/2}\right]$ and $\left[P^{-T/2}_{k|k}\hat x_{k|k}\right]$ that are directly read-off from the corresponding {\it extended}  post-array $[{\mathbb R}\: | \: \tilde b]$. This trick is intended to avoid any matrix inversion in the underlying filter recursion. In particular, in~\cite[Algorithm~2]{2017:CSL:Kulikova}  the Cholesky factor $R_{e,k}^{-1/2}$ is required for calculating $\hat x_{k|k}$, meanwhile the extended version in~\cite[Algorithm~3]{2017:CSL:Kulikova} does not involve it. Readers are referred to~\cite{2017:CSL:Kulikova} for more details and proof. Here we would like to discuss the possibility to design such methods for other factored-form MCC KF-like estimators. Our first question is whether or not the {\it extended} array implementations are practically feasible for the original MCC-KF recursion (Algorithm~1). We answer negatively for this question, because as mentioned in Section~\ref{SR:filters}, the equations for computing $K_k$ and $P_{k|k}$ in Algorithm~1 are taken from two different sources: the error covariance $P_{k|k}$ is computed by the classical KF equation (Joseph stabilized form), meanwhile the filter gain $K_k$ is calculated under the MCC methodology with implicated $\lambda_k$ parameter. In summary, these formulas have difference nature and cannot be collected altogether into unique array that is a crucial point for derivation of {\it extended} array implementations. Thus, the sign ``--'' is mentioned in the last row of Table~\ref{tab:properties} for all factored-form MCC-KF variants. Meanwhile for the IMCC-KF (Algorithm~2) recursion the extended array algorithms seems to be possible to derive. The Cholesky-based method has been recently suggested in~\cite[Algorithm~3]{2017:CSL:Kulikova}. The question about existence of extended array UD- and SVD-based IMCC-KF implementations is still open. This can be an area for future research.

\subsection{Numerical comparison}

To justify the theoretical derivation of the suggested factored-form implementations, a linear stochastic state-space model for electrocardiogram signal processing~\cite{2017:Suotsalo} is  explored. In contrast to the cited paper, the filtering methods are examined in the presence of impulsive noise/shot noise~\cite{2016:Izanloo}.

\begin{table*}
\caption{The RMSE errors and average CPU time (s) for the MCC-KF and IMCC-KF implementations in Example 1, $M = 500$ Monte Carlo simulations.} \label{tab:well}
\centering
{\small
\begin{tabular}{r||r|rrrr||r|rrr}
\hline
 & {\bf MCC-KF} & \multicolumn{4}{c||}{\bf Factored-form family for MCC-KF} & {\bf IMCC-KF}  & \multicolumn{3}{c}{\bf Factored-form family for IMCC-KF}\\
\hhline{*{2}{|~}*{4}{|-}|~|*{3}{|-}}
 & ({\it conventional}) & Cholesky- & UD- & SVD- & SVD- & ({\it conventional})  & Cholesky- & UD- & SVD-\\
 & Algorithm~1 & (1a) & (1b) & (1c) & (1d) & Algorithm~2 & \cite[Alg.~2]{2017:CSL:Kulikova} & (2b) & (2c) \\
\hline
RMSE$_{x_1}$  & 7.4691  & 7.4691  & 7.4691 & 7.4691 & 7.4691       & 7.3569  & 7.3569  & 7.3569  & 7.3569 \\
RMSE$_{x_2}$  & 12.4615  & 12.4615  & 12.4615 & 12.4615 & 12.4615  & 12.3938 & 12.3938 & 12.3938 & 12.3938 \\
RMSE$_{x_3}$  & 13.8206  & 13.8206  & 13.8206 & 13.8206 & 13.8206  & 13.7631 & 13.7631 & 13.7631 & 13.7631\\
\hline
$\|\mbox{\rm RMSE}_{x_i}\|_2$ & 20.0521  & 20.0521  & 20.0521 & 20.0521 & 20.0521  &  19.9287 & 19.9287 & 19.9287 & 19.9287 \\
CPU (s) & 0.0344  & 0.0543  & 0.0567 & 0.0680 & 0.0624  & 0.0264  & 0.0435  & 0.0472 & 0.0566 \\
\hline
\end{tabular}
}
\end{table*}

\begin{exmp} \label{ex:1}
 The system state is defined as $x(t) = [s(t), \: \dot{s}(t), \: \ddot{s}(t)]^T$ where $s(t)$ is the displacement of the object
or signal at time $t$, the derivatives $\dot{s}(t)$ and $\ddot{s}(t)$ represent the velocity
and acceleration, respectively. The discrete-time version of the model dynamic is given as follows:
\begin{align*}
x_{k} & =
\begin{bmatrix}
1 & \Delta t & \frac{(\Delta t)^2}{2} \\
0 & 1 & \Delta t \\
0 & 0 & 1
\end{bmatrix}\!\!
x_{k-1} + w_{k-1}, & x_0 & \sim {\cal N}(\bar x_0, \Pi_0)
\end{align*}
where $\Delta t = 0.1$ and $\bar x_0 = [1, 0.1, 0]^T$, $\Pi_0 = 0.1\:I_3$. The dynamic is observed via
the measurement scheme
\begin{align*}
y_k & =
\begin{bmatrix}
1 & 0 & 0
\end{bmatrix}
x_k + v_k.
\end{align*}
The entries of $w_k$ and $v_k$ are generated as follows:
\begin{align*}
w_k  & \sim {\cal N}(0, Q)+\mbox{\tt Shot noise}, \\
v_k  & \sim {\cal N}(0, R)+\mbox{\tt Shot noise}
\end{align*}
where the covariances $Q$ and $R$ are
\[
Q =
\begin{bmatrix}
\frac{1}{20}(\Delta t)^5 & \frac{1}{8}(\Delta t)^4 & \frac{1}{6}(\Delta t)^3 \\
\frac{1}{8}(\Delta t)^4 & \frac{1}{3}(\Delta t)^3 & \frac{1}{2}(\Delta t)^2 \\
\frac{1}{6}(\Delta t)^3 & \frac{1}{2}(\Delta t)^2 & \Delta t
\end{bmatrix} \quad  \mbox{and} \quad
R = 0.01.
\]
\end{exmp}

To simulate the impulsive noise (shot noise), we follow the approach suggested in~\cite{2016:Izanloo}. The Matlab routine \verb"Shot_noise" recently published in~\cite[Appendix]{2018:AJC:Kulikova} can be used as follows: (i) only 10\% of samples are corrupted by the outliers; (ii) the discrete time instants $t_k$ corrupted by the outliers are selected randomly from the uniform discrete distribution in the interval $[11,N-1]$, i.e. the first ten and last time instants are not corrupted in our experiments; (iii) the outliers are all taken at different time instants; (iv) the magnitude of each impulse is chosen randomly from the uniform discrete distribution in the interval $[0,3]$. Following~\cite{2016:Izanloo}, our routine additionally returns the sample covariances $\hat Q$ and $\hat R$ of the simulated random sequence. They are utilized by all estimators under examination.

To decide about estimation quality of each filtering method, the following numerical experiment is performed for $500$ Monte Carlo runs: (1) the stochastic model is simulated for $N=300$ discrete-time points to generate the measurements, (2) the inverse problem (i.e. the estimation problem) is solved by various filtering methods with the same measurement history, the same initial conditions, the same adaptive kernel size selection approach published previously for the MCC-KF method in~\cite{2016:Izanloo} and the same noises' covariances; (3) the root mean square error (RMSE) is calculated over $500$ Monte Carlo runs as follows:
\begin{equation*} \label{eq:RMSEx}
\mbox{\rm RMSE}_{x_i}=\sqrt{\frac{1}{MN}\sum \limits_{j=1}^{M} \sum \limits_{k=1}^N \left(x_{i, exact}^j(t_k) - \hat x_{i, k|k}^j\right)^2}
\end{equation*}
where $M=500$ is the number of Monte-Carlo trials, $N=300$ is the discrete time of the dynamic system, the $x_{i, exact}^j(t_k)$ and $\hat x_{i, k|k}^j$ are the $i$-th entry of the ``true'' state vector (simulated) and its estimated value obtained in the $j$-th Monte Carlo run, respectively. The resulted $\|\mbox{\rm RMSE}_{x_i}\|_2$ values are summarized in Table~\ref{tab:well} for each implementation under assessment. The averaged CPU time (s) is also presented for each estimator.

Having analyzed the obtained results collected at the first panel, we conclude  that all MCC-KF implementation methods derived in this paper are mathematically equivalent, i.e. the correctness of their derivation is substantiated by numerical experiments. The same conclusion holds for all IMCC-KF algorithms. Next, we observe that the SVD-based implementations are the most time consuming methods. As a benefit, we may mention that SVD provides an extra information about the matrix structure and, hence, these implementations might be used in various reduced-filters design strategies. The conventional algorithms are the most fast implementations, however, they are the most numerically unstable in ill-conditioned situations as we observe it in Example~2 below. Finally, having compared the results in the first and the second panels, we note that the IMCC-KF estimator (and all its factored-form implementations) outperforms the MCC-KF (and all its factored-form implementations, as well) for estimation accuracy. Indeed, the total RMSE of the IMCC-KF is less than the total RMSE of the MCC-KF method. The difference between them is caused by the neglected scaling parameter $\lambda_k$ in equation~\eqref{mcc:P:eq3}, i.e. this is the price to be paid for keeping the symmetric Joseph stabilized formula for $P_{k|k}$ calculation in the MCC-KF estimator.

Unfortunately, Example~1 does not allow for exploring numerical insights of the examined implementations. To do so, a set of ill-conditioned test problems is considered in Example~2.

\begin{exmp} \label{ex:1}
The dynamic equation in Example~1 is observed via
the following ill-conditioned scheme:
\begin{align*}
y_k & =
\begin{bmatrix}
1 & 1  & 1\\
1 & 1  & 1+\delta
\end{bmatrix}
x_k + v_k, \quad R \sim {\cal N}(0, \delta^2 I_2)
\end{align*}
with the initial state $x_0 \sim {\cal N}(0,I_3)$ and in the presence of Gaussian uncertainties, only. Additionally,
the ill-conditioning parameter $\delta$ is used for simulating roundoff and assumed to be
$\delta^2<\epsilon_{roundoff}$, but $\delta>\epsilon_{roundoff}$
where $\epsilon_{roundoff}$ denotes the unit roundoff
error\footnote{Computer roundoff for floating-point arithmetic is
often characterized by a single parameter $\epsilon_{roundoff}$,
defined as the largest number such that either
$1+\epsilon_{roundoff} = 1$ or $1+\epsilon_{roundoff}/2 = 1$ in
machine precision. }.
\end{exmp}

\begin{table*}
\caption{The effect of roundoff errors on the factored-form implementations designed for the MCC-KF (Algorithm~1) and IMCC-KF (Algorithm~2) estimators.} \label{tab:numeric}
\centering
{\small
\begin{tabular}{r||c|cccc||c|ccc}
\hline
Ill-conditioning &  {\bf MCC-KF} & \multicolumn{4}{c||}{\bf Factored-form family  for MCC-KF} & {\bf IMCC-KF}  & \multicolumn{3}{c}{\bf Factored-form family for IMCC-KF}\\
\hhline{*{2}{|~}*{4}{|-}|~|*{3}{|-}}
parameter &  ({\it conventional}) & Cholesky- & UD- & SVD- & SVD-  & ({\it conventional})  & Cholesky- & UD- & SVD-\\
$\delta$  &   Algorithm~1 & (1a) & (1b) & (1c) & (1d) & Algorithm~2 & \cite[Alg.~2]{2017:CSL:Kulikova} & (2b) & (2c) \\
\hline
$10^{-1}$  &    0.1192  & 0.1192 & 0.1192  & 0.1192  & 0.1192  & 0.1209  & 0.1209 & 0.1209 & 0.1209\\
$10^{-2}$  &    0.1000  & 0.1000 & 0.1000  & 0.1000  & 0.1000  & 0.1023  & 0.1023 & 0.1023 & 0.1023\\
$10^{-3}$  &    0.1040  &  0.1040 &  0.1040 &  0.1040 & 0.1040 & 0.1066 & 0.1066 & 0.1066 & 0.1066 \\
$10^{-4}$  &    0.1049 &  0.1049 &  0.1049 &  0.1049 & 0.1049 & 0.1069 & 0.1069 & 0.1069 & 0.1069 \\
$10^{-5}$  &    0.1018 & 0.1018 & 0.1018 & 0.1018 & 0.1018 & 0.1047 &  0.1047 &  0.1047 & 0.1047 \\
$10^{-6}$  &    0.0981 & 0.0981 & 0.0981 & 0.0981 & 0.0981 & 0.1008 &  0.1008 &  0.1008 &  0.1008 \\
$10^{-7}$  &    0.0997 & 0.0996 & 0.0996 & 0.0997 & 0.0997 & 0.1025 & 0.1026 & 0.1026 & 0.1029 \\
$10^{-8}$  &   \verb"NaN" & \verb"NaN" & \verb"NaN" & \verb"NaN" & 0.1016 & \verb"NaN" & 0.1046 & 0.1046 &  72.7067 \\
$10^{-9}$  &   \verb"NaN" & \verb"NaN" & \verb"NaN" & \verb"NaN" & 0.1004 & \verb"NaN" & 0.1032  & 0.1032  & \verb"Inf" \\
$10^{-10}$  &  \verb"NaN" & \verb"NaN" & \verb"NaN" & \verb"NaN" & 0.0915 & \verb"NaN" &  0.0936 &  0.0937  & \verb"NaN" \\
$10^{-11}$  &  \verb"NaN" & \verb"NaN" & \verb"NaN" & \verb"NaN" & 0.0868 & \verb"NaN" &  0.0890 & 0.0788 & \verb"NaN" \\
$10^{-12}$  &  \verb"NaN" & \verb"NaN" & \verb"NaN" & \verb"NaN" & 0.0997 & \verb"NaN" &  0.1024 & 0.1021 & \verb"NaN" \\
$10^{-13}$  &  \verb"NaN" & \verb"NaN" & \verb"NaN" & \verb"NaN" & 0.1003 & \verb"NaN" &  0.1027 &  0.1026 & \verb"NaN" \\
$10^{-14}$  &  \verb"NaN" & \verb"NaN" & \verb"NaN" & \verb"NaN" & 0.0985 & \verb"NaN" &   0.1010 &   0.1009  & \verb"NaN" \\
$10^{-15}$  &   \verb"NaN" & \verb"NaN" & \verb"NaN" & \verb"NaN" & 0.2341  & \verb"NaN" &  0.1012  & 0.1011  & \verb"NaN" \\
\hline
\end{tabular}
}
\end{table*}

The set of numerical experiments described above for Example~1 is performed for Example~2 as well, except that the covariance matrix of measurement noise $R$ remains the same, i.e. the process covariance $Q$ is replaced by the sample covariance $\hat Q$, only. The resulted $\|\mbox{\rm RMSE}_{x_i}\|_2$ values are summarized in Table~\ref{tab:numeric} for each implementation under assessment and each value of parameter $\delta$ while it tends to machine precision limit.

Having analyzed the numerical results collected in Table~\ref{tab:numeric}, we conclude that all filters produce accurate estimates of the state vector  while the estimation problem is well-conditioned, i.e. for large values of $\delta$. The resulted accuracy of the MCC-KF and IMCC-KF techniques is quite similar, until the problem becomes ill-conditioned and all implementations start to diverge. Recall, the difference in the original MCC-KF and the IMCC-KF is in the equation for $P_{k|k}$, only.

It is important to compare the factored-form implementations within the MCC-KF and IMCC-KF techniques, separately. For large $\delta$, we observe that the {\it factored-form} algorithms produce absolutely the same results compared with their {\it conventional} counterparts in Algorithm~1 or~2, respectively. Again, this substantiates the algebraic equivalence between the suggested factored-form implementations and the corresponding conventional algorithms. While $\delta$ tends to machine precision limit, some numerical insights can be explored. More precisely, starting from $\delta = 10^{-7}$ and less, the factored-form implementations behave in different manner. The SVD-based Algorithms~1c and~2c suggested in this paper produce a slightly less accurate estimates than the Cholesky- and UD-based implementations until their divergence at $\delta = 10^{-8}$. This outcome was expected and discussed in details in previous section. Recall, both Algorithms~1c and~2c imply the scalar divisions by square roots of the same singular values; see the terms $D_{P_{k|k-1}}^{-1/2}$ and $D_{P_{k|k}}^{-1/2}$ involved. Thus, their numerical behaviour is similar. Meanwhile, among all suggested SVD-based implementations, Algorithms~1d is the most robust (with respect to roundoff errors) and  it was anticipated in the previous section, as well. Indeed, the SVD-based MCC-KF implementation in Algorithm~1d maintains similar estimation accuracy as all other {\it factored-form} IMCC-KF implementations, i.e. the Cholesky-based Algorithm~1a and the UD-based Algorithm~1b. We also conclude that the SVD-based implementation (Algorithm~1d) is the only one method in the {\it MCC-KF factored-form} family that manages the examined ill-conditioned situations in Example~2. In contrast, the Cholesky-based Algorithm~1a and UD-based Algorithm~1b are the most robust implementations in the {\it factored-form} family derived for the IMCC-KF estimator. Recall, the question whether or not it is possible to design similar robust SVD-based IMCC-KF variant is still open and to be investigated in future. As discussed in previous section, a non-symmetric form of equation~\eqref{mcc:P:eq3} in the IMCC-KF estimator prevents the derivation.

Finally, we remark that all conventional implementations diverge at $\delta = 10^{-8}$. This conclusion holds for both the MCC-KF (Algorithm~1) and IMCC-KF (Algorithm~2). The term \verb"NaN" in Table~\ref{tab:numeric} means that the estimator cannot solve the filtering problem since it produces no correct digits in the obtained state vector estimate. Furthermore, the obtained numerical results demonstrate divergence of all factored-form implementations of the MCC-KF (Algorithm~1), except the SVD-based variant in Algorithm~1d. Meanwhile, we observe an accurate solution produced by the Cholesky- and UD-based implementations of the IMCC-KF (Algorithm~2). In total, there are only three implementations that manage the ill-conditioned state estimation problem while $\delta \to \epsilon_{roundoff}$. In summary, the family of robust {\it factored-form} MCC-KF implementations (with respect to roundoff errors) consists of only SVD-based method in Algorithm~1d, while the robust IMCC-KF implementations are the Cholesky- and UD-based Algorithms~2a and 2b.

\section{Concluding remarks}
In this paper, complete families of the factored-form implementations are derived for both the MCC-KF and the IMCC-KF estimators. The theoretical discussion and the results of numerical experiments indicate that only Cholesky- and UD-based IMCC-KF implementations solve the ill-conditioned state estimation problem accurately. For the MCC-KF estimator, the robust SVD-based implementation exists and only this algorithm accurately treats the ill-conditioned cases.

A number of questions are still open for a future research. First, for the MCC-KF estimator, only one robust implementation was found in the factored-form family of reliable algorithms. This is the SVD-based implementation. Thus, the proposed Cholesky- and UD-based MCC-KF implementations are to be improved in future research, if possible. In contrast, for the IMCC-KF estimator, the robust Cholesky- and UD-based implementation are derived in the factored-form family. Meanwhile, the derivation of robust SVD-based implementation for the IMCC-KF estimator is still an open question. Recall, the problem is how to balance the equation for error covariance matrix calculation in order to derive a symmetric form that is similar to the Joseph stabilized equation proposed for the classical Kalman filter. Next, all algorithms derived in this paper are of covariance type. Meanwhile, the information filtering under the MCC approach still does not exist, i.e. neither the conventional recursion nor the factored-form implementations have been derived, yet. The adaptive kernel size selection strategy is another importance problem for all MCC KF-like filtering methods. The small kernel size might induce the instability problem, as well. Hence, the related stability issues (with respect to kernel size selection) should be investigated. Furthermore, the extended array implementations are of special interests for a future research, because of improved numerical stability caused by utilization of stable orthogonal rotations as far as possible. Finally, the derivation of stable factored-form implementations for solving nonlinear  filtering problem under the maximum correntropy criterion via the accurate extended continuous-discrete KF approach presented recently in~\cite{KuKu16IEEE_TSP,KuKu16SISCI,KuKu17ANM,KuKu17SP,KuKu18IET_CTA} is also planned for a future research.

\section*{Acknowledgments}
The author thanks the support of Portuguese National Fund ({\it Funda\c{c}\~{a}o para a
Ci\^{e}ncia e a Tecnologia}) within the scope of project UID/Multi/04621/2013.

\section*{References}
\bibliographystyle{elsarticle-num}
\bibliography{BibTex_Library/books,%
              BibTex_Library/list_MVKulikova,%
              BibTex_Library/list_Tsyganova,%
              BibTex_Library/list_identification,%
              BibTex_Library/filters,%
              BibTex_Library/list_ML,%
              BibTex_Library/SVD,%
              BibTex_Library/MCCKF_Riccati,%
              BibTex_Library/list_linalg}

\begin{thebibliography}{10}
\expandafter\ifx\csname url\endcsname\relax
  \def\url#1{\texttt{#1}}\fi
\expandafter\ifx\csname urlprefix\endcsname\relax\def\urlprefix{URL }\fi
\expandafter\ifx\csname href\endcsname\relax
  \def\href#1#2{#2} \def\path#1{#1}\fi

\bibitem{2017:Aravkin}
A.~Aravkin, J.~V. Burke, L.~Ljung, A.~Lozano, G.~Pillonetto, {Generalized
  Kalman smoothing: Modeling and algorithms}, Automatica 86 (2017) 63--86.

\bibitem{1977:Masreliez}
C.~Masreliez, R.~Martin, {Robust Bayesian estimation for the linear model and
  robustifying the Kalman filter}, IEEE transactions on Automatic Control
  22~(3) (1977) 361--371.

\bibitem{2010:Hajiyev}
C.~Hajiyev, H.~E. Soken, {Robust estimation of UAV dynamics in the presence of
  measurement faults}, Journal of Aerospace Engineering 25~(1) (2010) 80--89.

\bibitem{2013:Chang}
L.~Chang, B.~Hu, G.~Chang, A.~Li, {Robust derivative-free Kalman filter based
  on Huber's M-estimation methodology}, Journal of Process Control 23~(10)
  (2013) 1555--1561.

\bibitem{2012:Charandabi}
B.~A. Charandabi, H.~J. Marquez, {Observer design for discrete-time linear
  systems with unknown disturbances}, in: IEEE 51st Annual Conference on
  Decision and Control, 2012, pp. 2563--2568.

\bibitem{2014:Charandabi}
B.~A. Charandabi, H.~J. Marquez, {A novel approach to unknown input filter
  design for discrete-time linear systems}, Automatica 50~(11) (2014)
  2835--2839.

\bibitem{2007:Liu}
W.~Liu, P.~P. Pokharel, J.~C. Pr{\'\i}ncipe, {Correntropy: properties and
  applications in non-Gaussian signal processing}, IEEE Transactions on Signal
  Processing 55~(11) (2007) 5286--5298.

\bibitem{2012:Cinar}
G.~T. Cinar, J.~C. Pr{\'\i}ncipe, {Hidden state estimation using the
  Correntropy filter with fixed point update and adaptive kernel size}, in: The
  2012 International Joint Conference on Neural Networks (IJCNN), 2012, pp.
  1--6.

\bibitem{2014:Chen}
B.~Chen, L.~Xing, J.~Liang, N.~Zheng, J.~C. Pr{\'\i}ncipe, {Steady-state
  mean-square error analysis for adaptive filtering under the maximum
  correntropy criterion}, IEEE Signal Processing Letters 21~(7) (2014)
  880--884.

\bibitem{2015:Chen}
B.~Chen, J.~Wang, H.~Zhao, N.~Zheng, J.~C. Pr{\'\i}ncipe, Convergence of a
  fixed-point algorithm under maximum correntropy criterion, IEEE Signal
  Processing Letters 22~(10) (2015) 1723--1727.

\bibitem{2016:Izanloo}
R.~Izanloo, S.~A. Fakoorian, H.~S. Yazdi, D.~Simon, {Kalman filtering based on
  the maximum correntropy criterion in the presence of non-Gaussian noise}, in:
  2016 Annual Conference on Information Science and Systems (CISS), 2016, pp.
  500--505.

\bibitem{2017:Chen}
B.~Chen, X.~Liu, H.~Zhao, J.~C. Pr{\'\i}ncipe, {Maximum Correntropy Kalman
  Filter}, Automatica 76 (2017) 70--77.

\bibitem{2018:Kulikov:SP}
G.~Y. Kulikov, M.~V. Kulikova, {Estimation of maneuvering target in the
  presence of non-Gaussian noise: A coordinated turn case study}, Signal
  Processing 145 (2018) 241--257.

\bibitem{2017:Liu:UKF}
X.~Liu, B.~Chen, B.~Xu, Z.~Wu, P.~Honeine, {Maximum correntropy unscented
  filter}, International Journal of Systems Science 48~(8) (2017) 1607--1615.

\bibitem{2017:Qin}
W.~Qin, X.~Wang, N.~Cui, {Maximum correntropy sparse Gauss-Hermite quadrature
  filter and its application in tracking ballistic missile}, IET Radar, Sonar
  \& Navigation 11~(9) (2017) 1388--1396.

\bibitem{2018:Principe:book}
D.~Comminiello, J.~C. Pr{\'\i}ncipe, Adaptive Learning Methods for Nonlinear
  System Modeling, 1st Edition, Elsevier, 2018.

\bibitem{2012:Chen}
B.~Chen, J.~C. Pr{\'\i}ncipe, {Maximum correntropy estimation is a smoothed MAP
  estimation}, IEEE Signal Processing Letters 19~(8) (2012) 491--494.

\bibitem{simon2006optimal}
D.~Simon, {Optimal state estimation: Kalman, H-infinity, and nonlinear
  approaches}, John Wiley \& Sons, 2006.

\bibitem{2017:CSL:Kulikova}
M.~V. Kulikova, {Square-root algorithms for maximum correntropy estimation of
  linear discrete-time systems in presence of non-Gaussian noise}, Systems \&
  Control Letters 108 (2017) 8--15.

\bibitem{GrewalAndrews2015}
M.~Grewal, A.~Andrews, Kalman filtering: theory and practice using MATLAB, 4th
  Edition, John Wiley \& Sons, New Jersey, 2015.

\bibitem{2010:Grewal:IEEE}
M.~S. Grewal, J.~Kain, {Kalman filter implementation with improved numerical
  properties}, IEEE Transactions on Automatic Control 55~(9) (2010) 2058--2068.

\bibitem{Morf1975}
M.~Morf, T.~Kailath, Square-root algorithms for least-squares estimation, IEEE
  Trans. Automat. Contr. AC-20~(4) (1975) 487--497.

\bibitem{Sayed1994}
A.~H. Sayed, T.~Kailath, Extended $\mathrm{C}$handrasekhar recursion, IEEE
  Trans. Automat. Contr. AC-39~(3) (1994) 619--622.

\bibitem{1995:ParkKailath}
P.~Park, T.~Kailath, New square-root algorithms for $\mathrm{K}$alman
  filtering, IEEE Trans. Automat. Contr. 40~(5) (1995) 895--899.

\bibitem{Bierman1977}
G.~J. Bierman, Factorization Methods For Discrete Sequential Estimation,
  Academic Press, New York, 1977.

\bibitem{2017:IET:KulikovaTsyganova}
M.~V. Kulikova, J.~V. Tsyganova, {Improved discrete-time Kalman filtering
  within singular value decomposition}, IET Control Theory \& Applications
  11~(15) (2017) 2412--2418.

\bibitem{1976:Thornton}
C.~L. Thornton, {Triangular covariance factorizations for Kalman filtering},
  Ph.D. thesis, University of California at Los Angeles (1976).

\bibitem{bierman1977numerical}
G.~Bierman, C.~Thornton, {Numerical comparison of Kalman filter algorithms:
  Orbit determination case study}, Automatica 13~(1) (1977) 23--35.

\bibitem{carlson1973fast}
N.~A. Carlson, Fast triangular formulation of the square root filter, AIAA
  journal 11~(9) (1973) 1259--1265.

\bibitem{Bjorck1967}
A.~Bj\"orck\, Solving least squares problems by orthogonalization, BIT 7 (1967)
  1--21.

\bibitem{1991:Gotze}
J.~G{\"{o}}tze, U.~Schwiegelshohn, {A square root and division free Givens
  rotation for solving least squares problems on systolic arrays}, SIAM Journal
  on Scientific and Statistical Computing 12~(4) (1991) 800--807.

\bibitem{1993:Hsieh}
S.~F. Hsieh, K.~J.~R. Liu, K.~Yao, {A unified square-root-free approach for
  QRD-based recursive-least-squares estimation}, IEEE Transactions on Signal
  Processing 41~(3) (1993) 1405--1409.

\bibitem{1994:Bjorck}
A.~Bj{\"{o}}rck.

\bibitem{JoverKailathSayed1986}
J.~M. Jover, T.~Kailath, A parallel architecture for kalman filter measurement
  update and parameter estimation, Automatica 22~(1) (1986) 43--57.

\bibitem{1991:Chun}
S.~H. Chun, {A single-chip QR decomposition processor for extended square-root
  Kalman filters}, in: The 25-th Asilomar Conference on Signals, Systems and
  Computers, 1991, pp. 521--529.

\bibitem{2017:Semushin}
I.~V. Semushin, Y.~V. Tsyganova, A.~V. Tsyganov, E.~F. Prokhorova, {Numerically
  efficient UD filter based channel estimation for OFDM wireless communication
  technology}, Procedia Engineering 201 (2017) 726--735.

\bibitem{WangSVD1992}
L.~Wang, G.~Libert, P.~Manneback, {Kalman Filter Algorithm based on Singular
  Value Decomposition}, in: Proceedings of the 31st Conference on Decision and
  Control, IEEE, Tuczon, AZ, USA, 1992, pp. 1224--1229.

\bibitem{Bjorck2015}
A.~Bj\"{o}rck, Numerical methods in matrix computations, Springer, 2015.

\bibitem{Dyer1969}
P.~Dyer, S.~McReynolds, Extensions of square root filtering to include process
  noise, J. Opt. Theory Appl. 3~(6) (1969) 444--459.

\bibitem{Golub1983}
G.~H. Golub, C.~F. Van~Loan, Matrix computations, Johns Hopkins University
  Press, Baltimore, Maryland, 1983.

\bibitem{Higham1990}
N.~J. Higham, {Analysis of the Cholesky decomposition of a semi-definite
  matrix}, Tech. Rep. MIMS EPrint: 2008.56, University of Manchester (1990).

\bibitem{2017:Suotsalo}
K.~Suotsalo, S.~S{\"a}rkk{\"a}, {A linear stochastic state space model for
  electrocardiograms}, in: The 27th IEEE International Workshop on Machine
  Learning for Signal Processing (MLSP), 2017.

\bibitem{2018:AJC:Kulikova}
M.~V. Kulikova, {Sequential maximum correntropy Kalman filtering}, Asian
  Journal of Control 21~(6) (2019) 1--9.

\bibitem{1965:Rauch}
H.~E. Rauch, C.~T. Striebel, F.~Tung, Maximum likelihood estimates of linear
  dynamic systems, AIAA journal 3~(8) (1965) 1445--1450.

\bibitem{KuKu16IEEE_TSP}
G.~Yu. Kulikov, M.~V. Kulikova, The accurate continuous-discrete extended
  \textsc\textit{{K}}alman filter for radar tracking, IEEE Trans. Signal
  Process. 64~(4) (2016) 948--958.

\bibitem{KuKu16SISCI}
G.~Yu. Kulikov, M.~V. Kulikova, Estimating the state in stiff continuous-time
  stochastic systems within extended \textsc\textit{{K}}alman filtering, SIAM
  J. Sci. Comput. 38~(6) (2016) A3565--A3588.

\bibitem{KuKu17ANM}
G.~Yu. Kulikov, M.~V. Kulikova, Accurate cubature and extended
  \textsc\textit{{K}}alman filtering methods for estimating continuous-time
  nonlinear stochastic systems with discrete measurements, Appl. Numer. Math.
  111 (2017) 260--275.

\bibitem{KuKu17SP}
G.~Yu. Kulikov, M.~V. Kulikova, Accurate continuous-discrete unscented
  \textsc\textit{{K}}alman filtering for estimation of nonlinear
  continuous-time stochastic models in radar tracking, Signal Process. 139
  (2017) 25--35.

\bibitem{KuKu18IET_CTA}
G.~Yu. Kulikov, M.~V. Kulikova,
  \textsc\textit{{M}}oore-\textsc\textit{{P}}enrose-pseudo-inverse-based
  \textsc\textit{{K}}alman-like filtering methods for estimation of stiff
  continuous-discrete stochastic systems with ill-conditioned measurementst,
  IET Control Theory Appl. 12~(16) (2018) 2205--2212.

\end{thebibliography}

\end{document}